\documentclass[prd,aps,preprint,amsmath,nofootinbib,amssymb,eqsecnum,showkeys,tightenlines]{revtex4-1}
\pdfoutput=1
\usepackage{tabularx}
\usepackage{slashed}
\usepackage{epsfig,latexsym,cancel,amssymb,amsmath,verbatim,mathrsfs}
\usepackage{color}
\usepackage{graphicx}
\usepackage{bm}
\usepackage{tabularx}
\usepackage{amsmath}
\usepackage{amssymb}
\usepackage{amsfonts}
\usepackage{cases}
\usepackage{cancel}
\usepackage{hyperref}
\usepackage{ulem}
\usepackage{here}
\usepackage{color}
\usepackage{graphicx}
\usepackage{diagbox}
\usepackage{multirow}
\usepackage{slashed}
\usepackage{simpler-wick}
\usepackage{subfigure}
\usepackage{svg}
\usepackage[marginal]{footmisc}
\usepackage{feynmf}

\def\vec{\mathbf}

\def\veck{\vec{k}}

\def\wt{\widetilde}

\newcommand{\be}{\begin{equation}}
\newcommand{\ee}{\end{equation}}
\newcommand{\bea}{\begin{eqnarray}}
\newcommand{\eea}{\end{eqnarray}}
\newcommand{\ba}{\begin{array}}
\newcommand{\ea}{\end{array}}

\def\wt{\widetilde}

\long\def\symbolfootnote[#1]#2{\begingroup%
\def\thefootnote{\fnsymbol{footnote}}\footnote[#1]{#2}\endgroup}

\newcommand{\diracslash}[1]{#1\!\!\!/}

\newcommand{\beq}{\begin{equation}}
\newcommand{\eeq}{\end{equation}}

\newcommand{\mrmC}[1]{\textcolor{magenta}{[\bf MRM: {#1}]}}

\begin{document}

\title{Bubble wall velocity from Kadanoff-Baym equations: fluid dynamics and microscopic interactions}

\author{Michael J.~Ramsey-Musolf}%
\email{mjrm@sjtu.edu.cn, mjrm@physics.umass.edu}
\affiliation{Tsung-Dao Lee Institute and  School of Physics and Astronomy, Shanghai Jiao Tong University,
800 Lisuo Road, Shanghai, 200240 China}
\affiliation{Shanghai Key Laboratory for Particle Physics and Cosmology, 
Key Laboratory for Particle Astrophysics and Cosmology (MOE), 
Shanghai Jiao Tong University, Shanghai 200240, China}
\affiliation{On Leave from Amherst Center for Fundamental Interactions, Department of Physics,
University of Massachusetts Amherst, MA 01003, USA }
\affiliation{ Kellogg Radiation Laboratory, California Institute of Technology,
Pasadena, CA 91125 USA}

\author{Jiang Zhu}
\email{jackpotzhujiang@gmail.com}
\affiliation{Tsung-Dao Lee Institute and  School of Physics and Astronomy, Shanghai Jiao Tong University,
800 Lisuo Road, Shanghai, 200240 China}
\affiliation{Shanghai Key Laboratory for Particle Physics and Cosmology, 
Key Laboratory for Particle Astrophysics and Cosmology (MOE), 
Shanghai Jiao Tong University, Shanghai 200240, China}

\date{\today}

\begin{abstract}
We establish a first principles, systematic framework for determining the bubble wall velocity during a first order cosmological phase transition. This framework, based on non-local Kadanoff-Baym equations, incorporates both macroscopic fluid dynamics and microscopic interactions between the bubble wall and particles in the plasma. Previous studies have generally focused on one of these two sources of friction pressure that govern the wall velocity. As a precursor, we utilize background field quantum field theory to obtain the relevant local Boltzmann equations, from which we derive the forces associated with variation of particle masses across the bubble wall and the microscopic wall-particle interactions. We subsequently show how these equations emerge from the Kadanoff-Baym framework under various approximations. We apply this framework in the ballistic regime to compute the new friction force arising from the $2\rightarrow 2$ scattering processes in scalar field theory. We obtain a linear relationship between this force and the Lorentz factor $\gamma_w$ that would preclude runaway bubbles with such effects.


\end{abstract}

\pacs{12.60.Jv,  14.70.Pw,  95.35.+d}

\maketitle

\section{Introduction}\label{incon}
First-order cosmological phase transitions can dramatically affect the thermal history of the universe. In a purely Standard Model (SM) universe, the transitions associated with the strong and electroweak sectors are known to be smooth crossovers: the transition from the deconfined to confined phase of quantum chromodynamics (QCD) at zero baryon chemical potential~\cite{Gross:1980br, Aoki:2006we, Kang:2021epo, Kang:2022jbg}, and the electroweak symmetry-breaking (EWSB) transition in the presence of a 125 GeV Higgs boson~\cite{Kajantie:1996mn,Dine:1992wr,Ramsey-Musolf:2019lsf}. 
In the latter case, inclusion of physics beyond the Standard Model (BSM) can render the character of EWSB transition to be first order. The presence of such a first order electrowewak phase transition (FOEWPT) -- if sufficiently \lq\lq strong\rq\rq\, -- can provide the necessary preconditions for electroweak baryogenesis (EWBG)~\cite{Kuzmin:1985mm, Cohen:1993nk, Rubakov:1996vz, Dine:2003ax, Morrissey:2012db} and a source of potentially observable gravitational waves (GW)~\cite{Apreda:2001us, Leitao:2012tx, Gould:2019qek, Kang:2020jeg, Friedrich:2022cak}. In some cases, it may also be associated with generation of the dark matter relic density~\cite{Falkowski:2012fb, Hambye:2018qjv, Baker:2019ndr, Chway:2019kft, Azatov:2021ifm}. 

A FOEWPT proceeds through bubble nucleation, wherein bubbles of broken symmetry, associated with non-zero masses for the SM elementary particles, nucleate and expand in the unbroken (massless particle) background. The bubble wall expansion velocity, $v_w$, is a crucial parameter for determining both the viability of EWBG and the characteristics of the GW signal. In the latter instance, for a \lq\lq runaway bubble\rq\rq\, wherein the wall velocity never reaches a terminal velocity, the associated GW spectrum will differ significantly from those generated by bubbles reaching a terminal $v_w$~\cite{Caprini:2015zlo, Ellis:2019oqb, Ellis:2020awk}. Moreover, different values of $v_w$ may lead to different GW spectra~\cite{Kamionkowski:1993fg, Caprini:2009yp, Hindmarsh:2013xza, Jinno:2016vai, Hindmarsh:2020hop, Athron:2023xlk, Kang:2025nhe}. Similarly, the viability of EWBG can also depend significantly on the value of $v_w$~\cite{Laurent:2020gpg,Cline:2021iff, Cline:2021dkf}.
Thus, it is important to obtain the most theoretically robust computation of $v_w$ when assessing the implications of a FOEWPT for these two phenomena.

\begin{figure}[htbp] 
    \centering
    \includegraphics[width=0.45\textwidth]{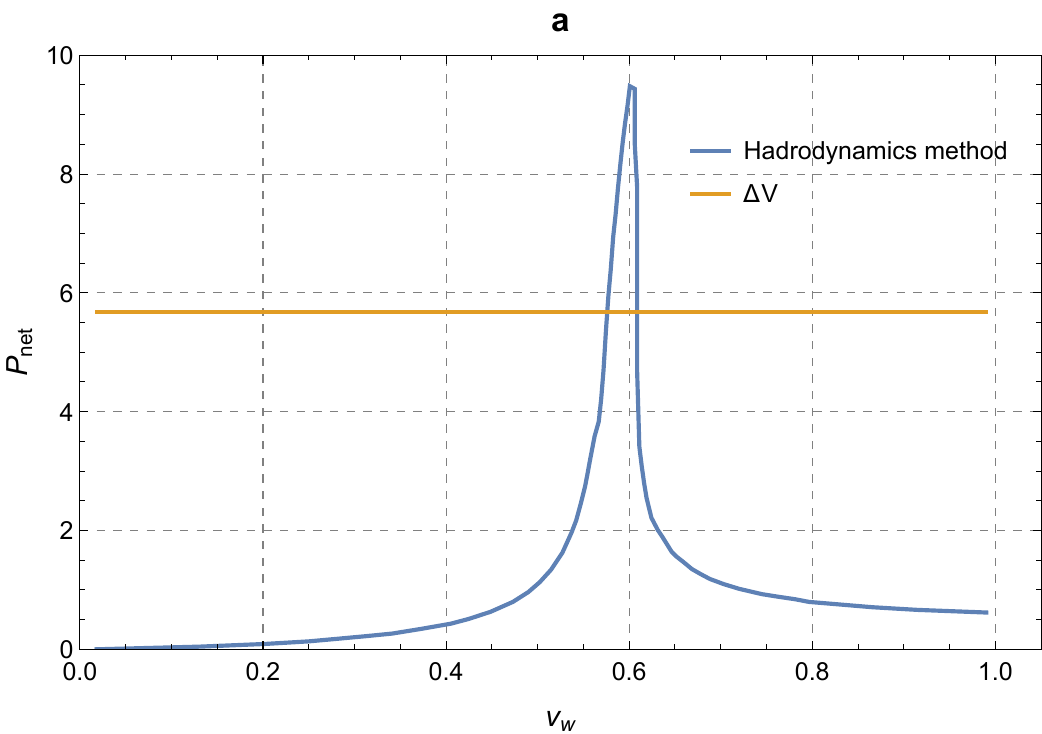} 
    \includegraphics[width=0.45\textwidth]{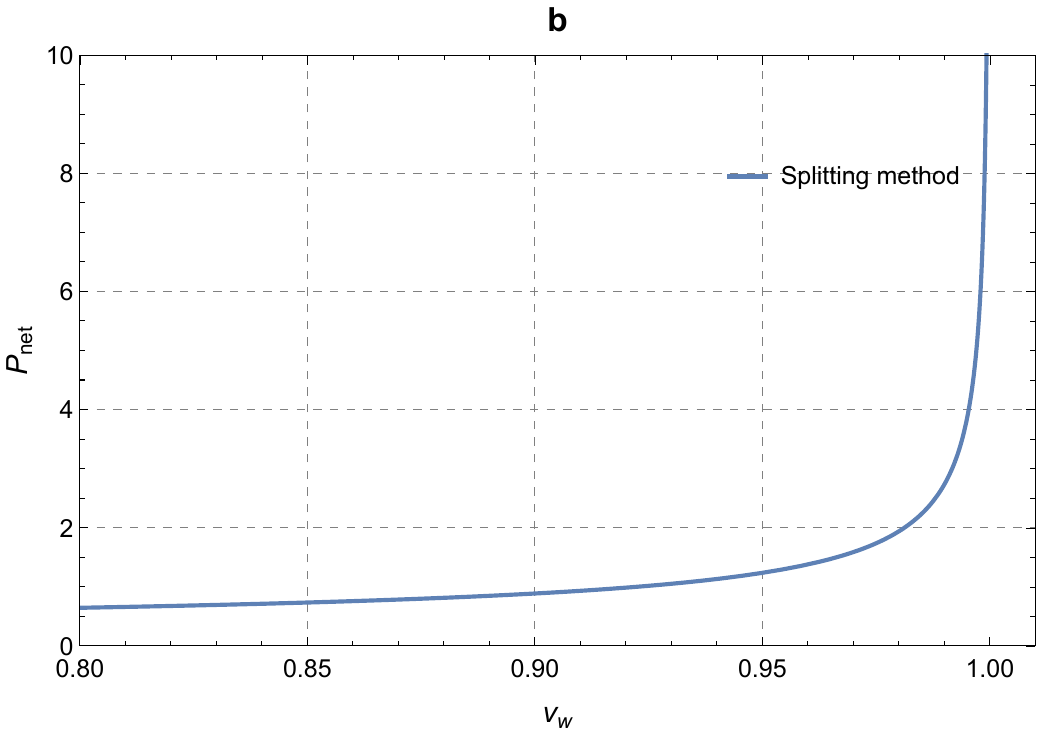} 
    \includegraphics[width=0.45\textwidth]{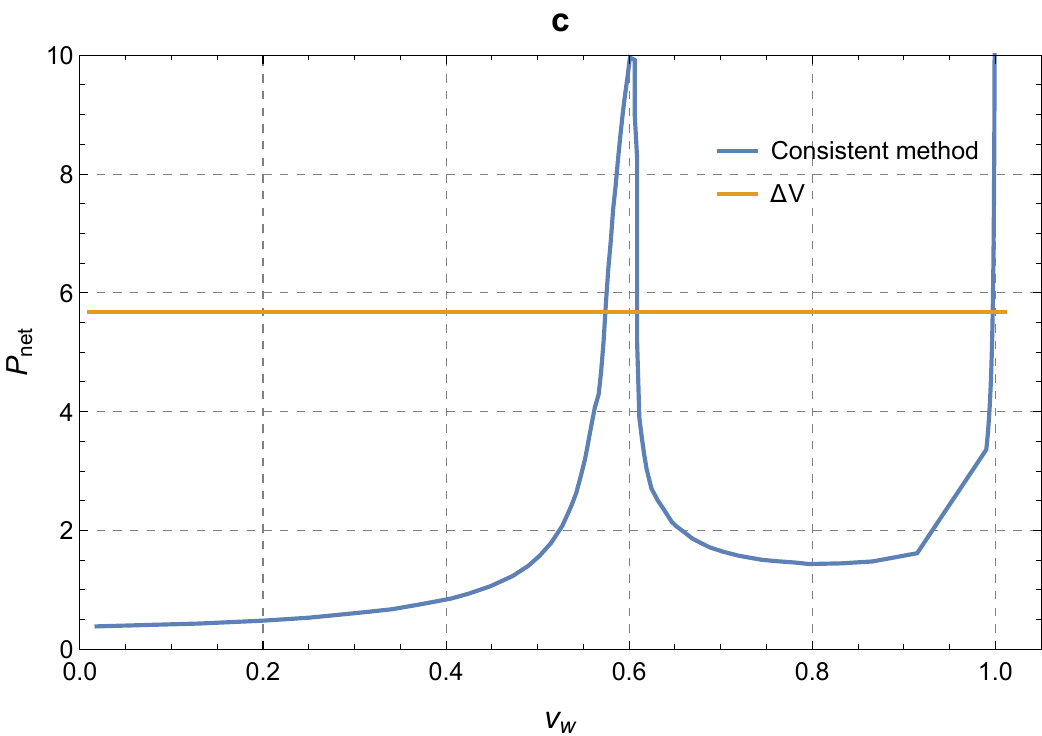} 
    \caption{
    Net pressure as a function of wall velocity $v_w$ for three cases: (a) Hydrodynamic contribution only (blue curve)~\cite{Laurent:2022jrs} (b) Splitting contribution only (blue curve)~\cite{Bodeker:2017cim} (c) expected total pressure come from the consistent treatment (hydrodynamic plus splitting) (blue curve). In panels (a) and (c) we show the corresponding driving pressure $\Delta V$ (brown line). Stable bubble wall velocity corresponds to blue and brown line intersection points.
    }
    \label{frictionFvsM}
\end{figure}

The key to computing this parameter is to consider the force balance between the driving force arising from the bubble interior-exterior vacuum energy difference and the plasma friction force. Two approaches have generally been employed to analyze this 
this force balance: the fluid method~\cite{Steinhardt:1981ct,Ignatius:1993qn, Moore:1995ua,Espinosa:2010hh,Konstandin:2014zta,Laurent:2022jrs,DeCurtis:2022llw,Ai:2023see,Li:2023xto,Wang:2023lam,DeCurtis:2022hlx} and the microscopic method~\cite{Arnold:1993wc,Bodeker:2009qy,Bodeker:2017cim,Azatov:2020ufh, Hoche:2020ysm,Azatov:2021ifm,Gouttenoire:2021kjv,Baldes:2022oev,Ai:2023suz,Azatov:2023xem,Kang:2024xqk}. The first method is widely used when 
the plasma satisfies the local thermal equilibrium condition~\cite{BarrosoMancha:2020fay,Ai:2021kak,Ai:2023see,Ai:2024shx}, and the second method is appropriate for the regime of ultra-relativistic wall velocity (ballistic approximation $v_w\sim 1$)~\cite{Dine:1992wr,Moore:1995si,Bodeker:2009qy}. However, these two methods are not consistent with each other. The friction force from the fluid method incorporates the contributions from the classical mass variation;  additional contributions are included by considering the multi-particle interaction processes in the microscopic method, which will eliminate the run-away bubble situation since the friction force is proportional to $\gamma_w$. This $\gamma_w$ dependence is induced by the Lorentz contraction of the particle density of the plasma $n_p$ since $F^\mathrm{fric}\propto n_p$. 
The mathematical description of this inconsistency can be found in the next Section. 
We illustrate the implications of these differences in Fig.~\ref{frictionFvsM}, where we plot the friction pressure 
as a function of $v_w$. Panel (a) illustrates the relationship arising from the fluid method, which manifests only one peak around the Joquet velocity, $v_J$.
Panel (b) illustrates the microscopic case, which yields a peak when $v_w\sim 1$. A fully consistent treatment should yield the two-peak structure of panel (c).

\begin{figure}[htbp] 
    \centering
    \includegraphics[width=0.4\textwidth]{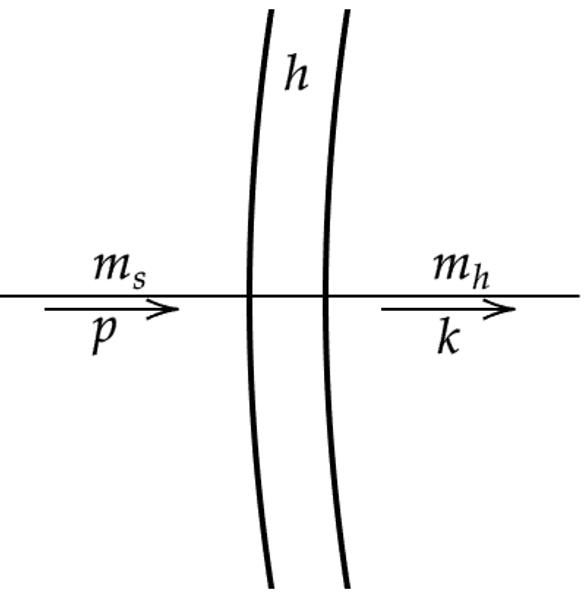} 
    \includegraphics[width=0.4\textwidth]{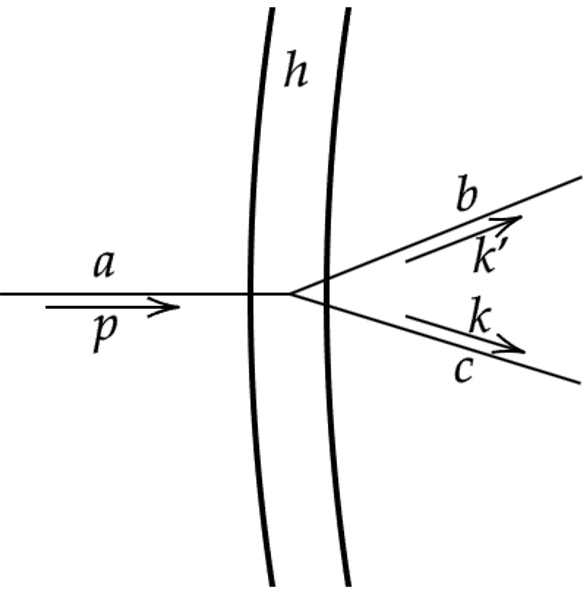} 
    \caption{The process of generating the friction pressure $F^{fric}_{i\rightarrow j}$. The left panel is the classical $1\rightarrow 1$ mass changing process; The right panel represents the multi-particle interaction, such as the $1\rightarrow 2$ splitting effect.}
    \label{micro}
\end{figure}

The inconsistency arises because in the fluid method the treatment of particle collisions 
always ensures four-momentum conservation~\cite{Calzetta:1986cq, Hohenegger:2008zk, Cirigliano:2009yt, Cirigliano:2011di,  Sheng:2021kfc, Ai:2023qnr}~\footnote{The momentum conservation in the ordinary Boltzmann equation must valid. This conclusion can be derived by the leading order Kadanoff-Baym equation.}, whereas in the microscopic method 
momentum non-conservation in the particle scattering process is considered when computing the friction pressure $F^\mathrm{fric}_{i\rightarrow j}/A$ in the $i\rightarrow j$ particle scattering process, which is demonstrated in Fig.\ref{micro}.
In this paper, we aim to resolve the above inconsistency by using quantum field theory in the presence of the background field of the bubble (BQFT). We subsequently show how the BQFT formulation emerges from the more general framework of the Kadanoff-Baym equations arising in the Closed Time Path (CTP) formulation of non-equilibrium quantum field theory. While BQFT is used in many other papers~\cite{Bodeker:2017cim,Gouttenoire:2021kjv,Azatov:2023xem,Kubota:2024wgx} we construct a systematic framework to compute the friction pressure for any multi-particle interaction process in the Boltzmann equation for the first time. We start with 
the kinetic evolution of particle distributions and derive the kinetic equation (Boltzmann equation) using the BQFT method. 
We start by discussing the quantization of the field with the BQFT method, then combining that quantization with the Boltzmann equation to obtain the friction pressure. 
In principle, using the modified Boltzmann equation, we can derive all possible friction forces in both the fluid and microscopic methods. We will prove the $F_{1\rightarrow 2}$ splitting friction force will indeed be generated by the collision term in the Boltzmann equation if we consider the BQFT. The friction forces from the multi-particle interactions will dominate when the bubble is extremely fast.

In addition, we show that the BQFT framework can be obtained starting from the Kadanoff-Baym equations, which provide the most general framework for treating non-equilibrium transport phenomena. We demonstrate that under certain well-motivated approximations the final result from the Kadanoff-Baym equation is, indeed, the Boltzmann equation obtained with the BQFT method. In this sense, we derive an ultimate equation for treating bubble wall velocity problems and the computation of $v_w$. In principle, by solving this equation, one can reconstruct both the friction generated by the fluid and the friction from multi-particle interaction and find all the information we need to analyze the dynamics governing the bubble wall velocity.

In what follows, we focus on the physics of the friction pressure and how to treat it self-consistently in the BQFT and Kadanoff-Baym frameworks. Beyond these considerations, there remains the practical challenge of computing $v_w$ starting from a physically complete and self-consistent quantum Boltzmann equation. In this work, we do not address this highly non-trivial practical challenge but refer the reader instead to recent work in Refs.~\cite{Laurent:2022jrs,DeCurtis:2022llw,DeCurtis:2023hil}. To connect our present analysis with these $v_w$ computations, we emphasize that the friction pressure is obtained as an integral over the collision terms $\mathcal{C}$ entering the Boltzmann equations, as illustrated in Eqs.~\eqref{eq:fricatobc} and \eqref{eq:fricitoj} below.
In all cases, a central point of our work is that the recent, widely-considered treatments of $v_w$ assume conservation of four-momentum in microscopic interactions. However, the contribution from the spacetime varying background field that contributes to elementary particle masses and, in some cases, the interaction vertices, breaks this conservation in the direction normal to the bubble wall ({\it e.g.}, $p_z$). Overall four momentum remains conserved when the background field contribution to the energy-momentum tensor is properly included, but the standard collision term integrals are modified by the non-conservation of microscopic $p_z$. An appropriate update of the state-of-art $v_w$ computations accounting for this effect is a task for future work.

Our discussion of these points in the remainder of the paper is organized as follows. In Section \ref{BQFT}, we systematically discussed the quantum field in the background field of the bubble wall and gave two commonly used approximations of the mode function. We briefly discuss the physics of the friction pressure from the Boltzmann equation for quantum mechanics in section.\ref{bleq} and prove that the collision term around the bubble wall is the origin of $1\rightarrow 2$ splitting friction pressure from the microscopic method. Then, we will solve the inconsistency in the same Section. In section.\ref{KBTOBE}, we will derive the modified Boltzmann equation from the Kadanoff-Baym equation and find that, under some approximation, these equations would generate the same friction pressure from the BQFT method. Using the collision terms around the bubble wall, we computed the friction pressure for $2\rightarrow 2$ and found it would linearly grow with the $\gamma_w$ of the bubble wall in section.\ref{22preV}. Conclusions and discussions, as well as the appendix, are cast in the remaining two sections.

\section{The Boltzmann equation in BQFT method}
\label{sec:BQFT}

The conventional Boltzmann equation (BE) treatment omits the presence of the spacetime varying background field and, thus, any impact of momentum non-conservation in microscopic interactions between particles in the plasma that manifests as non-local dynamics. To remedy this situation, we consider the non-local BE that incorporates the presence of the background field. To that end, we start with the most general form of the BE, whose derivation appears in Appendix \ref{bleq}:  


\begin{equation}\label{BE}
\begin{split}
    \frac{p^\mu}{E_p} \partial_\mu &f_a(x,p,t)+\partial_{p^i}[F^i f_a(x,p,t)]=
    \\
    &-f_a(x,p,t)\prod_{i\neq a}n_i\int\frac{\langle f|i\mathcal{T}|i\rangle\langle i|i\mathcal{T}|f\rangle}{T\prod_i (2E_i )V}\prod_{f}\frac{d^3\vec{p}_f[1\pm f_f(x,p_f,t)]}{2E_f(2\pi)^3}
    \\
    &+{\rm Inverse\ Process},
\end{split}
\end{equation}
Here, $f_a(x,p,t)$ is the phase space distribution function for particle species $a$ and $F^i=dp^i/dt$. The non-local effect of the background field enters the collision term through the amplitude $\langle f|iT|i\rangle$ when computed in the BQFT method. The latter has been adopted in earlier work~\cite{Bodeker:2017cim,Gouttenoire:2021kjv,Azatov:2023xem,Kubota:2024wgx} but, to our knowledge, has not been systematically applied in the BE context when computing $v_w$. We first summarize the BQFT framework, then apply it when utilizing Eq.~(\ref{BE}) to obtain the friction force balance needed to determine $v_w$.


\subsection{Quantum Field Theory Around the Bubble Wall}\label{BQFT}
We consider a bubble sufficiently large that one may neglect curvature (the planar approximation): the background field varies along the normal to the bubble wall, taken here to be the $z$-axis, but is homogeneous in the transverse directions. We further consider a plasma containing scalar, vector, and spinor fields, whose corresponding equations of motion (EOM) are 
\begin{equation}\label{EoMs}
\begin{split}
    &[\partial^2+m^2(z)]\phi(t,\vec{x}_\perp,z)=0
    \\
    &[\partial^2+m^2(z)]A_\mu(t,\vec{x}_\perp,z)=0
    \\
    &[i\diracslash{\partial}+m(z)]\psi(t,\vec{x}_\perp,z)=0.
\end{split}
\end{equation}
 where $\vec{x}_\perp$ indicates the coordinate perpendicular to the $z$ direction. We perform canonical quantization by utilizing the solutions to Eqs.~(\ref{EoMs}) as mode functions. 
We first consider the solution for scalar and vector fields with the similar EOM. Translation symmetry within the transverse bubble wall direction implies conservation of the perpendicular momentum $\vec{p}_\perp=(p_x,p_y)$. Thus, we can write the solution as a produce of a tansverse-direction plane wave and a normal direction mode function $\chi_k(z)$ as 
\begin{equation}\label{solutionform}
\begin{split}
    \phi_k(x)=\chi_k(z)e^{-i(Et-\vec{k}_\perp\cdot\vec{x}_\perp)},
\end{split}
\end{equation}
The solution for the vector field is the same but also includes an appropriate polarization vector. The normal direction mode function satisfies the Sturm-Liouvile type equation:
\begin{equation}\label{SLE}
    \frac{d^2\chi_k(z)}{dz^2}+k_z^2(z)\chi_k(z)=0,
\end{equation}
where $k_z^2(z)=E^2-|\vec{k}_\perp|^2-m^2(z)$.  
This is a Sturm-Liouville type equation, whose solutions one may use as a complete basis for expanding any function. The detailed behavior of these eigenfunctions depends, of course, on the $z$-dependence of $m^2(z)$. The authors of \cite{Azatov:2023xem} observe that a convenient choice of eigenfunctions are left- and right-moving modes, whose forms become linear combinations of transmitted and reflected plane waves asymptotically far from the wall.

The general solution of Eq.\eqref{SLE} depends on the form of the mass function $m(z)$. There are two limits for $m(z)$ that can lead to an analytic solution of Eq.\eqref{SLE}: 
\vskip 0.25in
\noindent (1) If $k_z\ll L_w^{-1}$, then one can use the step function approximation: $m(z)=m_s+(m_b-m_s)\Theta(z)$ In this case, one can formulate the solution in terms of products of the asymptotical plane wave  and the reflection and transition coefficients \cite{Azatov:2023xem}:
\begin{equation}\label{modethin}
\begin{split}
    \chi_{R,k}(z)=\begin{cases}
            e^{i k^z_s z} +  r  e^{-i k^z_s z}  \ , \\
            t_R e^{i k^z_b z}  \ , &
    \end{cases}
    \\
    \chi_{L,k}(z)=   \begin{cases}
           t _Le^{-i k^z_s z}  \ , \\
           r  e^{i k^z_b z}  +  e^{-i k^z_b z} \ , & 
    \end{cases}
\end{split}    
\end{equation}
where
\begin{equation}
    {k_z^{s,b}}=\sqrt{E^2-|\vec{k}_\perp|^2-m^2_{s/b}},\quad r=\bigg|\frac{k_z^s-k_z^b}{k_z^s+k_z^b}\bigg|\quad t_{R/L}=\bigg|\frac{2k_z^{s/b}}{k_z^s+k_z^b}\bigg|
\end{equation}
(2) On the other hand, for sufficiently large incident particle energy such that $k_z\gg L_w^{-1}$  one can imagine each spatial slice of the wall acts as a step function with negligible momentum change. In this regime,  $t/r \gg1$. Moreover,  for $k_z\gg L_w^{-1}$ or $k_z^2\gg k_z/L_w$ one has that $k_z(z)\gg k_z'(z)/k_z(z)$. In this case, one may adopt the WKB approximation~\footnote{As discussed in the next section, use of the WKB approximation  can help us to expand the quantum-kinetic equation in powers of the small quantity $1/(L_wk_z)$.} for these modes as

\begin{equation}\label{mode}
\begin{split}
    \chi_k(z)=\sqrt{\bigg|\frac{k_z^s}{k_z(z)}\bigg|}\exp\left({i\int_0^zdz'k_z(z')}\right),
\end{split}    
\end{equation}
where 
\begin{equation}\label{eq:kz}
k_z=\pm \sqrt{E^2-|\vec{k}_\perp|^2-m^2(z)}.
\end{equation}
The $k_z>0$ ($k_z<0$) branch represents the right-moving (left-moving) wave. 
In this work, where we consider right-moving 
particles incident on an ultra fast-moving wall, 
contributions from the reflected and left-moving modes are subdominant for distinct reasons. For a general phase transition with sufficiently large $v_w$, the left moving wave mode is subdominant since its contribution is exponentially suppressed by the distribution function $f\sim \exp[-\beta(E-\vec{k}\cdot\vec{v_w})]$ wherein $|E-\vec{k}\cdot\vec{v_w}| > E$ for $\vec{k}\cdot\vec{v_w}<0$.

The argument for neglecting the reflected mode is more subtle. If the average value of $\langle k_z\rangle\sim\gamma T>>\Delta m$, one has
\begin{equation}
 r=\bigg|\frac{k_z^s-k_z^b}{k_z^s+k_z^b}\bigg|\sim \frac{\Delta m}{\gamma T}<< t \sim 1 \ \ \ ,
\end{equation}
thereby allowing one to replace $r\to 0$ and $t\to 1$ in Eqs~(\ref{modethin}).
However, if the nucleation temperature is much smaller than $\Delta m$ such that the Lorentz factor cannot compensate, then the reflected mode should also be considered in the right-moving mode. We will not consider this situation in the remainder of this paper. The corresponding mode expansion for the scalar field operator then takes on the simple form
\begin{equation}\label{mode}
\hat{\phi}\approx\int\frac{d^3\vec{k}}{(2\pi)^3}\frac{1}{\sqrt{2E_k}}[\hat{a}_k\chi_k(z)e^{-i(Et-k_\perp\cdot x_\perp)}+\hat{a}^\dagger_k\chi_k^{*}(z)e^{i(Et-k_\perp\cdot x_\perp)}]
\end{equation}
$[a_k,a_p^\dagger]=(2\pi)^3\delta^3(k-p)$ and $[a_k,a_p]=[a_k^\dagger,a_p^\dagger]=0$. We note that in principle the $z$-component momentum in phase space integration is actually the eigenvalue in the Sturm-Liouville equation Eq.(\ref{SLE}), which means the $d^3\vec{k}\rightarrow d\lambda d^2\vec{k}_\perp$ with the eigenvalue $\lambda^2=E^2-\vec{k}_\perp^2$. However, if the WKB condition is satisfied $E^2-\vec{k}_\perp^2\approx k_z^2$, then we can approximately rewrite $d\lambda d^2\vec{k}_\perp$ as $d^3\vec{k}$. A more detailed discussion of this canonical quantization, including the corresponding formulation for spinor and vector fields, can be found in Appendix.\ref{quantization}.


After the quantization, we can look ahead and observe that the momentum non-conservation of the $z$-direction will emerge by employing Eq.\eqref{mode} to compute the matrix element appearing in Eq.\eqref{BE}. We also need to notice that the mode function in Eq.\eqref{mode} describes the particle which is asymptotically massless at $z\rightarrow-\infty$ but massive when $z\rightarrow\infty$.

\subsection{From Boltzmann equation to friction force}\label{sec:BtoF}
We now proceed to resolve he inconsistency problem in the context of the Boltzmann equation and BQFT.
First, multiply Eq.~(\ref{BE}) by $p^\nu$ and integrate $\int d^3 \vec{p}/(2\pi)^3$, leading to
\begin{equation}
    \int\frac{d^3\vec{p}}{(2\pi)^3}p^\nu\left[\frac{p^\mu }{E_p} \partial_\mu f_a+\partial_{p^i}(F^i f_a)\right]=\int\frac{d^3\vec{p}}{(2\pi)^3}p^\nu\mathcal{C}[f_a]
\end{equation}
where the collision term $\mathcal{C}[f_a]$ is given by
\begin{equation}\label{colc}
    \mathcal{C}[f_a]=-f_a(x,p,t)\prod_{i\neq a}n_i\int\frac{\langle f|i\mathcal{T}|i\rangle\langle i|i\mathcal{T}|f\rangle}{T\prod_i (2E_i )V}\prod_{f}\frac{d^3\vec{p}_f[1\pm f_f(x,p_f,t)]}{2E_f(2\pi)^3}+{\rm Inverse\ Process},
\end{equation}
For the particle \lq\lq $a$\rq\rq\, passing through the wall, the force is given by 
$\vec{F}=-\vec{\nabla} m_a^2/(2E_p)$, where the spatial variation of $m_a^2$ arises from the contribution to the mass from the scalar background field that defines the bubble. Integrating the second term by parts gives
\begin{equation}\label{eq:bzbz}
    \partial_\mu\int\frac{d^3\vec{p}}{(2\pi)^3}\frac{p^\mu p^\nu}{E_p}  f_a=\int\frac{d^3\vec{p}}{(2\pi)^3}F^\nu f_a+\int\frac{d^3\vec{p}}{(2\pi)^3}p^\nu\mathcal{C}[f_a]
\end{equation}
where we have ignored the surface term and taken $F^\mu=(0,\vec{F})$. We consider a bubble sufficiently large to allow us to use planar approximation and take the normal to the wall to be the $z$-axis. In addition, since we are only interested in the terminal bubble wall velocity that is defined by the stable bubble wall, any derivative with time will vanish. In this case, $\partial_\mu f_a$ and $F^\nu$ are zero except for $\mu,\nu=z$. Thus
\begin{equation}\label{eq:BE2}
    \partial_z\int\frac{d^3\vec{p}}{(2\pi)^3}\frac{p_z^2}{E_p}  f_a=\int\frac{d^3\vec{p}}{(2\pi)^3}F^z f_a+\int\frac{d^3\vec{p}}{(2\pi)^3}p^z\mathcal{C}[f_a]
\end{equation}
The statistical definition of the energy-momentum tensor for particles of species $a$,  is
\begin{equation}
    T_a^{\mu\nu}=\int\frac{d^3p}{(2\pi)^3}\frac{p^\mu p^\nu}{E_p}f_a(p)\ \ \ ,
\end{equation}
which has the same units as pressure. Thus, we can identify that the LHS of Eq.~(\ref{eq:BE2}) is the rate of change pressure with $z$, which is a force per volume.
The first term on the RHS is the friction force density(force times the particle's number density) $F_{1\rightarrow 1}/V$ arising from the classical process of the mass variation of particle $a$, which is demonstrated in the first panel of Fig.\ref{micro}. 
As we will find in Eq.\eqref{motransfe} below, the second term in RHS is the momentum density transfer between the particle and bubbles in the collision process. 
As derived in Appendix~\ref{bleq}, the collision term fundamentally represents the product of scattering rates (events per unit time) and distribution functions. Through phase space integration, its final form manifests as momentum transfer per unit spacetime volume, equivalent to a force density.
This term is normally zero as long as the momentum is conserved in the microscopic process~\cite{Cirigliano:2009yt,Konstandin:2014zta, Hindmarsh:2020hop}. 

However, let us consider the scattering process in the presence of the background field. Due to the interaction between the particle and the spatial dependence of the background field, the momentum conservation in the multiple particle scattering process is broken.
For example, the splitting process shown in the right panel of Fig.\ref{micro}, generates a non-trivial contribution to the last term of Eq.\eqref{eq:BE2}.

More generally, for the background field-dependent process, the collision terms 
need not vanish 
since the momentum is not conserved in the $z$-direction. Consequently, the momentum-space interaction amplitudes carry an additional spatial dependence ($z$-dependence), in contrast to the usual quantum field theory computations. We will prove this conclusion later.
If the last term of Eq.\eqref{eq:BE2} is nonzero, one can treat it as a friction force density. To summarize: the first term of the RHS is the force density from the classical mass variation, and the second term is the force density from the background field-dependent scattering process. Therefore, we can find the RHS to be exactly the total friction force density from the particle $a$ given by the microscopic method. Thus, the Eq.\eqref{eq:BE2} is the equation that describes the equation that the force density should satisfy for the static bubble wall.

To obtain the pressure from the total friction force density, we only need to integrate this equation in the $z$-direction and sum up all particles:
\begin{equation}\label{Fric}
\begin{split}
    \sum_a \Delta\bigg[\int\frac{d^3\vec{p}}{(2\pi)^3}\frac{p_z^2}{E_p}  f_a\bigg]^{z=s}_{z=h}=\sum_a\int dz\int\frac{d^3\vec{p}}{(2\pi)^3}F^z f_a+\sum_a\int dz\int\frac{d^3\vec{p}}{(2\pi)^3}p^z\mathcal{C}[f_a],
\end{split}
\end{equation}
where $a$ in the summation index represents all particles in the fluid, index $s/h$ represents the position of the symmetry/broken vacuum~\footnote{In general, the position of symmetry/broken vacuum can be taken to be $-\infty/\infty$.}, and $\Delta[f(x)]^{x=a}_{x=b}=f(a)-f(b)$. Let us discuss the physical meaning of this equation. To this end, it is useful to rewrite Eq.~(\ref{Fric}) in terms of the back-reaction process and fluid energy-momentum tensor. The LHS as  
\begin{equation}\label{rbae}
\begin{split}
    \Delta\left[T^{zz}_f(z)\right]^{z=s}_{z=h}=-\frac{F^\mathrm{fric}_\mathrm{Bubble}}{A}.
\end{split}
\end{equation}
where we have identified  $F^\mathrm{fric}_\mathrm{Bubble}/A$ 
as the friction pressure acting on the bubble. The minus sign on the RHS of Eq.~(\ref{rbae}) is introduced because the force acting on the bubble has the same magnitude but opposite direction from the force acting on the particle (the RHS of Eq.\eqref{Fric}). The above equation tells us that the friction pressure provided by the bubble wall, which contains both classical force and force brought by the background field dependent scattering process, causes the energy-momentum change of fluid.


Now we demonstrate that the collision term of Eq.\eqref{eq:BE2} is the momentum transfer between the particle and bubble in the scattering process. We start from the definition of the momentum transfer by a microscopic process: the transfer of momentum flux current times the probability that generates those momentum transfer:
\begin{equation}\label{motransfe}
\begin{split}
    \frac{F^\mathrm{fric}_{a\rightarrow bc}}{A}&=\int\frac{d^3\vec{p}_a}{(2\pi)^3}\frac{p_a^z}{E_a}f_a\int dP_{a\rightarrow bc}(p_a^z-p_b^z-p_c^z),
\end{split}
\end{equation}
which is first shown in~\cite{Bodeker:2017cim}. From the quantization in Section.\ref{BQFT}, we can compute the probability $dP_{a\rightarrow bc}$ as 
\begin{equation}\label{prob}
\begin{split}
    \int dP_{a\rightarrow bc}&=\int\frac{d^3\vec{p}_b}{(2\pi)^3}\frac{1}{2E_b}\int\frac{d^3\vec{p}_c}{(2\pi)^3}\frac{1}{2E_c}\frac{1}{2p_a^z}|M|^2_{a\rightarrow bc}(2\pi)^3\delta(E_a-E_b-E_c)\delta^2(\vec{p}_a^\perp-\vec{p}^\perp_b-\vec{p}^\perp_c),
\end{split}
\end{equation}
where 
\begin{equation}\label{amplitude}
M_{a\rightarrow bc}=\int dz \chi_a^*(z)V(z)\chi_b(z)\chi_c(z)\approx\int dzV(z)e^{-i[p_z^a(z)-p_z^b(z)-p_z^c(z)]z}
\end{equation}
with $V(z)$ being the vertex function for the $a\rightarrow bc$ process~\cite{Bodeker:2017cim,Ai:2023suz,Azatov:2023xem}, the detialed derivation can be find in Eq.\eqref{E4}
. Then substituting Eq.\eqref{prob} into the definition Eq.\eqref{motransfe}, we obtain
\begin{equation}
\begin{split}\label{spliting}
    \frac{F^\mathrm{fric}_{a\rightarrow bc}}{A}=\int\frac{d^3\vec{p}_a}{(2\pi)^3}\frac{f_a}{2E_a}\int&\frac{d^3\vec{p}_b}{(2\pi)^3}\frac{d^3\vec{p}_c}{(2\pi)^3}\frac{|M|^2_{a\rightarrow bc}}{2E_b 2E_c}(p_a^z-p_b^z-p_c^z)
    \\
    &\times(2\pi)^3\delta(E_a-E_b-E_c)\delta^2(\vec{p}_a^\perp-\vec{p}^\perp_b-\vec{p}^\perp_c).
\end{split}
\end{equation}
The mathematical detail of this derivation can be found in the Appendix.\ref{splitting}.
Now let us prove the collision terms in Eq.(\ref{Fric}) have the same expression as shown in Eq.\eqref{motransfe} and Eq.\eqref{spliting}.

We first consider the splitting process $a\rightarrow bc$ in the last term of Eq.(\ref{Fric})
\begin{equation}
\begin{split}
    \int dz\int\frac{d^3p_a}{(2\pi)^3}p^z_a\mathcal{C}[f_a]=-\int dz\int\frac{d^3\vec{p}_a}{(2\pi)^3}p^z_af_a(p_a,x)\int\frac{|\langle b,c|i\mathcal{T}|a\rangle|^2}{2E_aVT}\frac{d^3\vec{p}_b}{2E_b(2\pi)^3}\frac{d^3\vec{p}_c}{2E_c(2\pi)^3}&
    \\
    +\int dz\int\frac{d^3\vec{p}_b}{(2\pi)^3}\int\frac{d^3\vec{p}_c}{(2\pi)^3}p^z_af_b(p_b,x)f_c(p_c,x)\int\frac{|\langle a|i\mathcal{T}|b,c\rangle|^2}{2E_b2E_cVT}\frac{d^3\vec{p}_a}{2E_a(2\pi)^3}&
\end{split}
\end{equation}
where we have omitted the final state statistical factor through $1+f\approx1$. The analogous equation for particle $b$
is given by
\begin{equation}\label{eq:ab}
\begin{split}
    \int dz\int\frac{d^3p_i}{(2\pi)^3}p^z_b\mathcal{C}[f_b]=&-\int dz\int\frac{d^3\vec{p}_b}{(2\pi)^3}\int\frac{d^3\vec{p}_c}{(2\pi)^3}p^z_b f_b(p_b,x)f_c(p_c,x)\int\frac{|\langle a|i\mathcal{T}|b,c\rangle|^2}{2E_b2E_cVT}\frac{d^3\vec{p}_a}{2E_a(2\pi)^3}
    \\
    &+\int dz\int\frac{d^3\vec{p}_a}{(2\pi)^3}p^z_b f_a(p_a,x)\int\frac{|\langle b,c|i\mathcal{T}|a\rangle|^2}{2E_aVT}\frac{d^3\vec{p}_b}{2E_b(2\pi)^3}\frac{d^3\vec{p}_c}{2E_c(2\pi)^3}
\end{split}
\end{equation}
while the equation for particle $c$ can be found by interchanging $b\leftrightarrow c$ in Eq.\eqref{eq:ab}, 
then use the WKB quantization in Section.\ref{BQFT} to evaluate the amplitude as outlined in detail in Appendix~\ref{splitting}:  
\begin{equation}\label{defamp}
    \langle b,c|iT|a\rangle=M_{a\rightarrow bc}(2\pi)^3\delta(E_a-E_b-E_c)\delta^{(2)}(\vec{p}_a^\perp-\vec{p}_b^\perp-\vec{p}_c^\perp)
\end{equation}
Thus, we obtain 
\begin{equation}
\begin{split}
    \sum_{i=a}^{b,c}\int dz\int\frac{d^3p_i}{(2\pi)^3}p^z_i\mathcal{C}[f_i]&=-\int dz\int\frac{d^3\vec{p}_a}{(2\pi)^3}f_a(p_a,x)\int\frac{d^3\vec{p}_b}{2E_b(2\pi)^3}\frac{d^3\vec{p}_c}{2E_c(2\pi)^3}\frac{|M_{a\rightarrow bc}|^2}{2E_aL}(2\pi)^3
    \\
    &\quad\quad\quad\quad\quad\times\delta(E_a-E_b-E_c)\delta^{(2)}(\vec{p}_a^\perp-\vec{k}_b^\perp-\vec{k}_c^\perp)(p^z_a-p^z_b-p^z_c)
    \\
    &+{\rm Inverse\ Process}.
\end{split}
\end{equation}
where the factor of $L$ comes from the the box normalization condition $2\pi\delta(0)=L$ and $V=L^3$. More detailed information can be found in Appendix~\ref{splitting}.

To achieve our goals, let us constrain our discussion later 
by two commonly used approximations for relativistic bubble wall velocity:
\begin{itemize}
    \item 
    Approximation-A: We assume that the mass variation of the particles is small compared with the energy of the incoming particles, $\Delta m/E_p\ll 1$, which allows us to use the WKB approximation of the mode function.
    \item Approximation-B: We treat the distribution function as the equilibrium distribution function for incoming particles $f(k,X)\approx f_{in}(k)$. This approximation is also used in the microscopic method and should be correct if the bubble wall velocity is large enough, since it is the ballistic solution of the Boltzmann equation~\footnote{In general, when the bubble wall speed is very large $\gamma_w\gg1$, the mean free path of the interaction between particles $L_{MFP}$ is much larger than the thickness of the bubble $L_w/\gamma_w$, then it is good approximation to ignore the collision terms in the Boltzmann equation. Thus, the solution of distribution is dominated by the Liouville terms in the Boltzmann equation~\cite{Arnold:1993wc,Moore:1995si,Moore:1995ua,Bodeker:2009qy,Ai:2024btx} and is called the \lq\lq ballistic approximation\rq\rq .}.
\end{itemize}
If the distribution functions for the incoming particles and $M_{a\rightarrow bc}$ are $z$-independent (Approximation-B){
, then the integral over $dz$ will contribute an overall factor of $L$, leading to 
\begin{equation}
\begin{split}
    \sum_{i=a}^{b,c}\int dz\int\frac{d^3p_i}{(2\pi)^3}p^z_i\mathcal{C}[f_i]&=-\int\frac{d^3\vec{p}_a}{(2\pi)^3}\frac{f_a(p_a)}{2E_a}\int\frac{d^3\vec{p}_b}{2E_b(2\pi)^3}\frac{d^3\vec{p}_c}{2E_c(2\pi)^3}\frac{|M_{a\rightarrow bc}|^2}{2E_a}(2\pi)^3
    \\
    &\quad\quad\quad\times\delta(E_a-E_b-E_c)\delta^{(2)}(\vec{p}_a^\perp-\vec{k}_b^\perp-\vec{k}_c^\perp)(p^z_a-p^z_b-p^z_c)
    \\
    &+{\rm Inverse\ Process}
\end{split}
\end{equation}
It is precisely the microscopic momentum transfer between bubble particles in Eq.(\ref{spliting}). 
Thus, we have shown that the collision terms in the Boltzmann equation are the origin of the microscopic splitting friction force 
\begin{equation}
\label{eq:fricatobc}
    \frac{F^\mathrm{fric}_{a\rightarrow bc}}{A}+{\rm Inverse\ Process}=-\sum_{i=a}^{b,c}\int dz \int\frac{d^3\vec{p}_i}{(2\pi)^3}p^z_i\mathcal{C}[f_i].
\end{equation}
The foregoing argument and be generalized for any
$i\rightarrow f$ scattering process as
\begin{equation}
\label{eq:fricitoj}
\begin{split}
    \frac{F^\mathrm{fric}_{i\rightarrow f}}{A}&+{\rm Inverse\ Process}
    \\
    &=-\sum_{a=1,..}^i\int dz \int\frac{d^3\vec{p}_a}{(2\pi)^3}p_a^z\mathcal{C}[f_a]-\sum_{b=1,...}^f\int dz\int\frac{d^3\vec{p}_b}{(2\pi)^3}p_b^z\mathcal{C}[f_b],
\end{split}
\end{equation}
and the friction force is written by the BQFT amplitude as:
\begin{equation}\label{fricGn}
\begin{split}
    \frac{F^\mathrm{fric}_{i\rightarrow f}}{A}&=\Pi_i\int\frac{d^3\vec{p}_i}{(2\pi)^3}\frac{f(p_i)}{2E_i}\Pi_f\int\frac{d^3\vec{k}_f}{(2\pi)^3}\frac{1}{2E_f}|M|^2_{i\rightarrow f}(\sum_{i}p_i^z-\sum_fk_f^z)
    \\
    &\quad\quad\quad\quad\quad\ \times(2\pi)^3\delta(\sum_iE_i-\sum_fE_f)\delta^{(2)}(\sum_i\vec{p}_i^\perp-\sum_f\vec{k}_f^\perp)
    \\
    &+{\rm Inverse\ Process}\ \ \ ,
\end{split}
\end{equation}
which gives the friction pressure arising from the $i\rightarrow j$ process. 

With the same derivation, we can also express Eq.\eqref{BE} into the more explicit expression with the $z$-direction propagating bubble wall and $1\rightarrow2$ splitting effect
\begin{equation}\label{BE:COMPARE}
\begin{split}
    \bigg[2k_z&\frac{\partial}{\partial z}-\frac{dm^2(z)}{dz}\frac{\partial}{\partial k_z} \bigg]\frac{f_a(k,z)}{E_k}=
    \\
    &-\int\frac{d^3\vec{p}_b}{(2\pi)^3}\int\frac{d^3\vec{p}_c}{(2\pi)^3}\frac{f_a(k,z)|M_{a\rightarrow bc}|^2}{E_aL}\frac{1\pm f_b(p_b,z)}{2E_b}\frac{1\pm f_c(p_c,z)}{2E_c}
    \\
    &\quad\quad\quad\times(2\pi)^3\delta(E_k-E_p-E_{p'})\delta^2(\vec{k}_\perp-\vec{p}_\perp-\vec{p}'_\perp)+{\rm Inverse\ Process}.
\end{split}
\end{equation}
where the detailed derivation can be found in Appendix.\ref{splitting}.

\subsection{Solving the Inconsistency}\label{IncS}
Now, let us address the inconsistency between the fluid and the microscopic method when computing the bubble wall velocity. In the previous section, we have shown that the non-trivial friction force from the microscopic method can be consistently treated by the Boltzmann equation with the BQFT method. Thus, one can expect that the inconsistency introduced in Section~\ref{incon} can be resolved by introducing the modified Boltzmann equation with the BQFT method.

To that end, we recall that the fluid method entails utilizing the Boltzmann equation plus the total energy-momentum conservation. For the bubble-plasma coupled system, the total energy momentum tensor can be written as $T^{\mu\nu}=T_f^{\mu\nu}+T_\phi^{\mu\nu}$. Here $T_f^{\mu\nu}$ is the plasma energy momentum tensor in the previous discussion, and $T^{\mu\nu}_\phi$ is the standard scalar field energy-momentum tensor, which describes the bubble
\begin{equation}
    T^{\mu\nu}_\phi=\partial^\mu\phi\partial^\nu\phi-g^{\mu\nu}\left[\frac{1}{2}(\partial\phi)^2-V(\phi)\right].
\end{equation}
To derive the fluid equations of motion (EOM) 
, the fluid method uses the ordinary Boltzmann equation, which assumes momentum conservation in the microscopic particle interactions at the bubble wall. Thus, one finds the same equation as Eq.\eqref{eq:bzbz} without the contribution from the collision terms
\begin{equation}\label{FEMT}
    \partial_\mu T_f^{\mu\nu}=\sum_a\int\frac{d^3\vec{p}}{(2\pi)^3}F^\nu f_a \ \ \ ,
\end{equation}
where $T^{\mu\nu}_f$is the fluid contribution to the energy momentum tensor.
Now, using $\partial_\mu T^{\mu\nu}_\mathrm{total}=0$, we can replace the LHS of Eq.\eqref{FEMT} by $\partial_\mu T_f^{\mu\nu}\rightarrow -\partial_\mu T_\phi^{\mu\nu}$. 
Recall that we seek to determine the terminal bubble wall velocity. In this case, the field configuration in the bubble wall reference frame is time independent. Integrating the equations along the direction normal to the bubble wall integral of the LHS  Eq.\eqref{FEMT}  is
\begin{equation}\label{eq:conser}
    \int dz \partial_zT^{z\nu}_f=-\int dz \partial_zT^{z\nu}_\phi=-\Delta T_{\phi}^{z\nu}\ \ \ .
\end{equation}
To get a more simplified expression, one should also notice that the field configuration is also coordinate-independent outside of the bubble wall.
Thus, any terms containing derivatives in the RHS of the  Eq.\eqref{eq:conser} 
will vanish. Taking $\nu=z$ the RHS of Eq.~\eqref{eq:conser} then becomes the difference of the potential 
\begin{equation}\label{condic}
    \Delta[V(\phi)]^{z=h}_{z=s}=\frac{F^\mathrm{fric}_{1\rightarrow 1}}{A}.
\end{equation}
Importantly, Eq.\eqref{condic} omits the collision effect. 

Now, in the microscopic method, the driving force provided by the vacuum energy difference should be equal to the sum of all the friction forces
\begin{equation}\label{bacond}
    \Delta[V(\phi)]^{z=h}_{z=s}=\frac{F^\mathrm{fric}_\mathrm{Bubble}}{A}=\frac{F^\mathrm{fric}_{1\rightarrow 1}}{A}+\sum_{i,j}\frac{F^\mathrm{fric}_{i\rightarrow j}}{A},
\end{equation}
in order to counter deriving pressure provided by the LHS of \eqref{bacond} and reach the force balance state. 
By comparing Eq.\eqref{condic} and Eq.\eqref{bacond}, it can be shown that an additional friction pressure term arises on the RHS of Eq.\eqref{bacond} due to the non-conservation of momentum in the z-direction. If the momentum non-conservation effect is not properly included, this extra friction term would vanish, leading to significant physical consequences. In systems governed predominantly by Eq.\eqref{condic}, bubbles would run-away. However, when the momentum non-conservation is incorporated self-consistently, the bubbles reach a terminal velocity. This, in turn, affects the fraction of the latent heat released by the phase transition that is deposited into the bubbles and the surrounding fluid. As a consequence, the difference would affect the GWs as an observational signal.


By introducing the non-trivial background field effect, the inconsistency between the fluid and microscopic methods can be immediately solved by replacing Eq.\eqref{FEMT} with Eq.\eqref{eq:BE2}. One find that  
\begin{equation}\label{BQFTBE}
\begin{split}
    \int dz\partial_z T_\phi^{z z}=\Delta[V(\phi)]^{z=h}_{z=s}&=\sum_a\int dz\int\frac{d^3\vec{p}}{(2\pi)^3}F_w^z f_a
    \\
    &+\sum_i\Pi_i\int\frac{d^3\vec{p}_i}{(2\pi)^3}\frac{f(p_i)}{2E_i}\Pi_f\int\frac{d^3\vec{k}_f}{(2\pi)^3}\frac{1}{2E_f}\frac{|M|^2_{i\rightarrow f}}{L_w}(\sum_{i}p_i^z-\sum_fk_f^z)
    \\
    &\quad\quad\quad\quad\quad\quad\quad\times(2\pi)^3\delta(\sum_iE_i-\sum_fE_f)\delta^{(2)}(\sum_i\vec{p}_i^\perp-\sum_f\vec{k}_f^\perp)
    \\
    &+{\rm Inverse\ Process}
    \\
    &=\frac{F_{1\rightarrow 1}}{A}+\sum_{i,j}\frac{F_{i\rightarrow j}}{A}.
\end{split}
\end{equation} 

By addressing the inconsistency between the two methods, we observed that the solution of the modified Boltzmann equation should, in principle, give us the friction force in Figure.\ref{frictionFvsM}. Because, as we proved in the previous section, if $v_w\sim1$, we can use the ballistic approximation and find the microscopic friction force that gives us a friction pressure as shown in the right upper panel of Figure.\ref{frictionFvsM}. However, for $v_w< 1$ where the heat conduction rate brought by the collision term is fast enough to generate the local thermal equilibrium in which the fluid configuration is nearly independent of the interaction~\footnote{More detailed discussion can be found in~\cite{Ai:2021kak}, which shown the friction pressure in first panel of Fig.\ref{frictionFvsM} can be directly obtained by consider the fluid configuration in the classical mass variation.}, then the friction pressure from the mass variation contribution would generate the result shown in the left upper panel of Figure.\ref{frictionFvsM}~\cite{Ai:2021kak,Ai:2023see,Ai:2024btx}. Combining the two branches, we find the friction force which forms just as the lower panel of Fig.\ref{frictionFvsM}.

\section{The Kadanoff-Baym Equation}\label{KBTOBE}

Although the above discussion has been able to solve the bubble wall velocity computation inconsistency problem, it remains to be shown how such a resolution follows from first principles in quantum field theory (QFT). Indeed, the treatment in Section \ref{sec:BtoF} utilizes BQFT to compute the relevant transition amplitudes, but they are embedded in the Botlzmann equation (BE) derived from classical kinetic theory. A fully self-consistent, first principles QFT approach requires use of out-of-equilibrium quantum field theory as formulated, for example, in the Schwinger-Keldysh formalism. In this context, the Kadanoff-Baym equations (KBEs) fully describe the evolution of particle densities in the presence of spacetime varying background fields ({\it e.g.}, the scalar field bubble wall profile), interactions, and thermal effects. Under certain assumptions and approximations, one may obtain the conventional Boltzmann equations, as utilized in Section \ref{sec:BtoF}. In this section, we show how the BQFT results in Section \ref{sec:BtoF} emerge from the KBEs under certain assumptions. 

Our approach is analogous to the treatment in Refs.~\cite{Cirigliano:2009yt,Cirigliano:2011di} that perform a systematic expansion of the KBEs in relevant scale ratios, denoted generically as $\epsilon$. The BE arises after expanding the constraint and kinetic KBEs to zeroth and first orders in $\epsilon$, respectively\footnote{The momentum non-conservation will appear in the non-local term in Kadanoff-Baym equation~\cite{Wagner:2022amr}}. Here, we focus on $\epsilon_\mathrm{wall}$ that characterizes gradients with respect to bubble wall. We peform an $\epsilon_\mathrm{wall}$ expansion everywhere except when gradients act on or within a $\delta$-function. In this context we show how the background field effect and momentum non-conservation normal to the wall at leading non-trivial order in wall gradients.  

\subsection{From Kadanoff-Baym Equation to Boltzmann Equation}\label{OKB-BE}
The KBE can be derived from the Schwinger-Dyson equation for the two-point Green's function $G(x,y)$ in the Closed Time Path (CTP) formalism (see {\it {e.g.}}, \cite{Das:1997gg} for more details). Let us rewrite $G(x,y)$ as a function of the average and different of coordinates $X=\frac{x+y}{2}$, $r=x-y$  and define the Wigner transformation of function as
\begin{equation}\label{wigner}
    G(k,X)=\int d^4r G(x,y)e^{ik\cdot r}\quad\quad\Pi(k,X)=\int d^4r \Pi(x,y)e^{ik\cdot r}.
\end{equation}
Taking the sum and differences of the Schwinger-Dyson equations and performing the Wigner transformation yields a pair of scalar field KBEs
\begin{equation}\label{KBE}
\begin{split}
    &(\frac{1}{2}\partial_X^2-2k^2)G^{\gtrless}+e^{-i\Diamond}\{m^2,G^{\gtrless}\}=-ie^{-i\Diamond}(\{\Pi^h,G^{\gtrless}\}+\{\Pi^{\gtrless},G^h\}+\frac{1}{2}[\Pi^{>},G^{<}]-\frac{1}{2}[G^{>},\Pi^{<}]),
    \\
    &-2ik\cdot \partial_X G^{\gtrless}+e^{-i\Diamond}[m^2,G^{\gtrless}]=-ie^{-i\Diamond}([\Pi^h,G^{\gtrless}]+[\Pi^{\gtrless},G^h]+\frac{1}{2}\{\Pi^{>},G^{<}\}-\frac{1}{2}\{G^{>},\Pi^{<}\}),
\end{split}
\end{equation}
where $G^{a}(k,X)$ and $\Pi^a(k,X)$ represent the Wigner transformed Greens function and self-energy, respectively, and where the  index $a=>,<,h$ indicates branches of the closed time path\footnote{The $G^\gtrless$ constitute the so-called Wightman functions}. The KBEs contain two sub-equations: the first is the constraint equation that determines the spectral content,  and the second is the kinetic equation that governs the dynamics.  Both equations contain the diamond operator $\Diamond$, defined by
\begin{equation}
\Diamond(A B)=\frac{1}{2}(\partial_X A\cdot\partial_k B-\partial_k A\cdot\partial_X B).
\end{equation}
The details about the derivation and the CTP formalism can be found in  Appendix ~\ref{driveofKBE}.

As a non-linear set of coupled, integral-differential equations, the KBEs are challenging to solve. In the present context, one may obtain an approximate solution by expanding in the aforementioned scale ratios. In particular, we follow Ref.~~\cite{Cirigliano:2009yt} and consider 
\begin{equation}
\begin{split}
    \epsilon_{wall}=\frac{\tau_{int}}{\tau_{wall}}\quad\quad\epsilon_{coll}=\frac{\tau_{int}}{\tau_{coll}},
\end{split}
\end{equation}
where $\tau_{int}\sim 1/\omega$, with $\omega$ being the quasiparticle frequency; $\tau_{wall}\sim \partial_X$; and $\tau_{coll}$ being the time scale associated with particle collisions in the plasma. At $\mathcal{O}(\epsilon^0)$ the constraint equation yields the following form for the Wightman functions: 
\begin{equation}\label{SCKBE}
\begin{split}
    G^>(k,X)=2\pi\delta[k^2-m(X)^2](\Theta(k^0)[1+f(k,X)]+\Theta(-k^0)\bar{f}(-k,X))
    \\
    G^<(k,X)=2\pi\delta[k^2-m(X)^2](\Theta(k^0)f(k,X)+\Theta(-k^0)[1+\bar{f}(-k,X)])
\end{split}
\end{equation}
where $f$ ($\bar{f}$) represents the particle (anti-particle) distribution. Substituting (\ref{SCKBE}) into the $\mathcal{O}(\epsilon)$ kinetic equation and integrating $\int_0^\infty \frac{dk^0}{(2\pi)^4}$ gives
\begin{equation}\label{LOKBE}
    \left[2k\cdot\partial_X -\vec{\nabla}_X m^2(X)\cdot \frac{\partial }{\partial\vec{k}}\right]\frac{f(k,X)}{E_k}=\int_0^\infty\frac{dk^0}{2\pi}\frac{1}{2}(\{\Pi^>,G^<\}-\{\Pi^<,G^>\}),
\end{equation}
If we consider the cuts in the 1-loop self-energy $\Pi^a$ then the RHS of Eq.(\ref{LOKBE}) will become collision term in Eq.(\ref{ncol}) at tree-level. In this case, Eq.(\ref{LOKBE}) is exactly the Boltzmann equation with collision term given by leading non-trivial order  of $e^{-i\Diamond}\{\Pi^a,G^b\}$. 

\subsection{The Modified Boltzmann equation from Complete Kadanoff-Baym Equation}\label{FrictioninKBE}
To obtain the background field-dependent effect in the Boltzmann equation and, ultimately, the associated non-conservation of momentum in the direction normal the wall, one has to go beyond the leading orders (in $\epsilon$-counting) of the KBE~\cite{Cirigliano:2009yt,Cirigliano:2011di}. In this case, one can naturally ask whether the systematic inclusion or higher-orders in the $\epsilon$-expansion
 generates the momentum non-conservation effect as seen in the BQFT calculation. We show below that doing so does not yield the expected effect.
In short,   if we want to obtain the friction pressure resulting from the momentum non-conservation  then we may not be able to derive relevant equation  by naive application of the $\epsilon_{wall/coll}$ expansion. Rather, one must take care to identify where the this expansion can be applied and where the resummation to all orders of $\epsilon_{wall/coll}$ is needed.

To our knowledge, there exists no systematic method to solve the complete KBE in full generality. 
Consequently, we will proceed by adopting three  approximations:
\begin{itemize}
    \item Approximation-1:  Perform the $\epsilon_{wall/coll}$ expansion and ignore the higher order terms except in places where there exists a singularity that precludes such an expansion
    \item Approximation-2: Formulate the Wightman function using the solution from the $\mathcal{O}(\epsilon^0) $ constraint equation. In principle, one may include modifications of the spectral content arising from high-order in $\epsilon$ terms in the constraint equation. In practice, it suffices to employ the LO constraint equation to make the connection with the previously discussed BQFT results.
    \item Approximation-3: The Approximation-B in Section \ref{sec:BQFT}. 
\end{itemize}
In fact, these three approximations are equivalent to the two approximations that we used in the section\ref{sec:BtoF}, and Approximation-2 can be derived from Approximation-1. However, we will defer a discussion of 
the physical meaning and rationality of Approximation-1 and Approximation-2 in the last part of this subsection. 

To proceed from the full kinetic equation, we first recall that the only spatial dependence is in the $z$-direction, so we retain only the $z$-derivatives in $\partial_X$. It is then convenient to integrate $\int_0^\infty\frac{dk^0}{2\pi}$, yielding

\begin{equation}\label{eq:gb}
\begin{split}
    \int_0^\infty\frac{dk^0}{2\pi}k_z\frac{d}{dz}G^{<}(k,z)&+\frac{i}{2}\int_0^\infty\frac{dk^0}{2\pi}e^{-i\Diamond}[m_a^2(z),G^{<}(k,z)]
    \\
    &= \frac{1}{4}\int_0^\infty\frac{dk^0}{2\pi}e^{-i\Diamond}(\{\Pi^>_a,G^{<}_a\}-\{\Pi^<_a,G^{>}_a\})
    \\&+\frac{1}{2}\int_0^\infty\frac{k^0}{2\pi}e^{-i\Diamond}([\Pi^h_a,G^{<}_a]+[\Pi^<_a,G^{h}_a]).
\end{split}
\end{equation}

Let us discuss this equation term by term. 

\vskip 0.1in
\noindent A. First term on the LHS: 
By using Eq.\eqref{SCKBE} according to Approximation-2, it is straightforward to obtain
\begin{equation}
    \int_0^\infty\frac{dk^0}{2\pi}k_z\frac{d}{dz}G^{<}(k,z)=k_z\frac{\partial}{\partial z}\frac{f(k,z)}{2E_k(z)}.
\end{equation}

\vskip 0.1in
\noindent B. The second term on the LHS contains the diamond operator. To get an analytical expression, we use the following property of the diamond operator  
\begin{equation}
e^{-i\Diamond}(A(k,x)B(k,x))=A(k+i\partial_x,x-i\partial_k)B(k,x)=B(k-i\partial_x,x+i\partial_k)A(k,x)
\label{eq:diamondidentity}
\end{equation}
and find that
\begin{eqnarray}\label{eq:seten}
    \int_0^\infty\frac{dk^0}{2\pi}k_z e^{-i\Diamond}[m_a^2(z),G^{<}(k,z)]&=&k_z \int_0^\infty\frac{dk^0}{2\pi}\bigg[m_a^2(z-\frac{id}{2dk_z})G^{<}(k,z)\\
\nonumber   
    &&-m_a^2(z+\frac{id}{2dk_z})G^{<}(k,z)\bigg].
\end{eqnarray}
As we mentioned above, we must be careful when performing the $\epsilon$-expansion to check whether there is a singularity. 
To that end, consider the formal Taylor expansion
\begin{equation}
m_a^2(z\mp \frac{id}{2dk_z}) = m_a^2(z) \mp  \frac{dm_a^2}{dz} \, \frac{id}{2dk_z})+\cdots = m_a^2(z) \mp
\mathcal{O}(\epsilon_{wall})\times\frac{i}{2}\frac{d}{dk_z}+\mathcal{O}(\epsilon_{wall}^2)\ \ \ .
\end{equation}
where the gradient $dm_a^2/dz$ is $\mathcal{O}(\epsilon_{wall}). $
In a naive application of the $\epsilon$-expansion, one would first retain all terms through $\mathcal{O}(\epsilon_{wall})$ and omit higher order terms. However, we observe from \eqref{SCKBE} that the spectral representation of the Greens function contains a $\delta$-function involving $k$. Nominally, the action of $d/d k_z$ on this $\delta$-function would introduce a singularity, thereby rendering the $\epsilon$-expansion ill-defined. Nevertheless, we can avoid this situation by first performing the $k^0$ integral, yielding
\begin{eqnarray}
\nonumber
    \int_0^\infty\frac{dk^0}{2\pi}k_z e^{-i\Diamond}[m_a^2(z),G^{<}(k,z)]&=&k_z\bigg(m_a^2(z-\frac{id}{2dk_z})\frac{f(k,z)}{2E_k(z)}-m_a^2(z+\frac{id}{2dk_z})\frac{f(k,z)}{2E_k(z)}\bigg)\\
    \label{eq:fided}
    &=& - i\frac{k_z}{2}\frac{dm_a^2(z)}{dz}\frac{\partial}{\partial k_z}\frac{f(k,z)}{E_k(z)}+\mathcal{O}(\epsilon^2_{wall})\ \ \ .
\end{eqnarray}

As a consequence, after the above discussion, Eq.\eqref{eq:gb} can now be written as 
\begin{equation}
    k_z\frac{\partial}{\partial z}\frac{f(k,z)}{2E_k(z)}+\frac{k_z}{2}\frac{dm_a^2(z)}{dz}\frac{\partial}{\partial k_z}\frac{f(k,z)}{E_k(z)}+\mathcal{O}(\epsilon^2_{wall/coll})=\mathcal{C}_1+\mathcal{C}_2
\end{equation}
where $\mathcal{C}_{1,2}$ indicate the contributions from the two terms on the RHS of Eq.~(\ref{eq:gb}), respectively. It is similar in structure to the ordinary Boltzmann equation: the first terms indicate the kinetic terms; the $\mathcal{O}(\epsilon_{wall})$ term containing $dm_a^2/dz$ 
corresponds to the semiclassical force term arising from the variation of the mass across the wall. We identify the remaining terms containing the self-energy functions as the collision terms.

\vskip 0.1in
\noindent C. For the first collision term $\mathcal{C}_1$, the treatment of the derivatives and $\delta$-functions is more subtle and results in the emergence of the $z$-momentum non-conservation effect.  To proceed, let us rewrite the $\mathcal{C}_1$ term as
\begin{equation}
\begin{split}\label{FPCKBE}
    \int_0^\infty\frac{dk^0}{2\pi}&k_ze^{-i\Diamond}\{\Pi^>(k,z),G^<(k,z)\}
    \\
    &=\int_0^\infty\frac{dk^0}{2\pi}k_z\bigg[\Pi^>(k+\frac{i}{2}d_z,z-\frac{i}{2}d_{k_z})G^<(k,z)+G^<(k+\frac{i}{2}d_z,z-\frac{i}{2}d_{k_z})\Pi^>(k,z)\bigg]
    \\
    &=\int_0^\infty\frac{dk^0}{2\pi}k_z\bigg[\Pi^>(k+\frac{i}{2}d_z,z-\frac{i}{2}d_{k_z})G^<(k,z)+\Pi^>(k-\frac{i}{2}d_z,z+\frac{i}{2}d_{k_z})G^<(k,z)\bigg].
\end{split}
\end{equation}
where we have used Eq.~(\ref{eq:diamondidentity}); $d^{i}_{z/k_z}$ represents the four vectors $d_{z/k_z}=(0,0,0,\partial_{z/k_z})$; and where we have assumed that the distribution function is only dependent on the coordinate $z$. The other part of the $\mathcal{C}_1$  can be obtained explicitly by exchange $>\leftrightarrow<$. One can easily prove this equation by performing a Taylor expansion. 

To continue the discussion, we must specify an interaction to write down the form of the self-energy $\Pi^>$. For the sake of simplicity and for purposes of illustration, we chose a toy model
\begin{equation}\label{lasp}
    \mathcal{L}_{int}=Y(z)\ \phi\ \Phi^2
\end{equation}
which has been discussed in previous literature in the context of the BQFT method~\cite{Azatov:2021ifm,Ai:2023suz}. Although we have assumed a specific interaction here, the discussion will not lose generality since other interactions can be discussed similarly. The friction pressure generated by this interaction comes from the simplest $1\rightarrow 2$ process $\phi\rightarrow \Phi\Phi$, illustrated in the right panel of Fig.\ref{micro}. The corresponding self-energy diagram is shown in Fig.\ref{SELFENERGY}.
\begin{figure}[htbp] 
    \centering
    \includegraphics[width=0.6\textwidth]{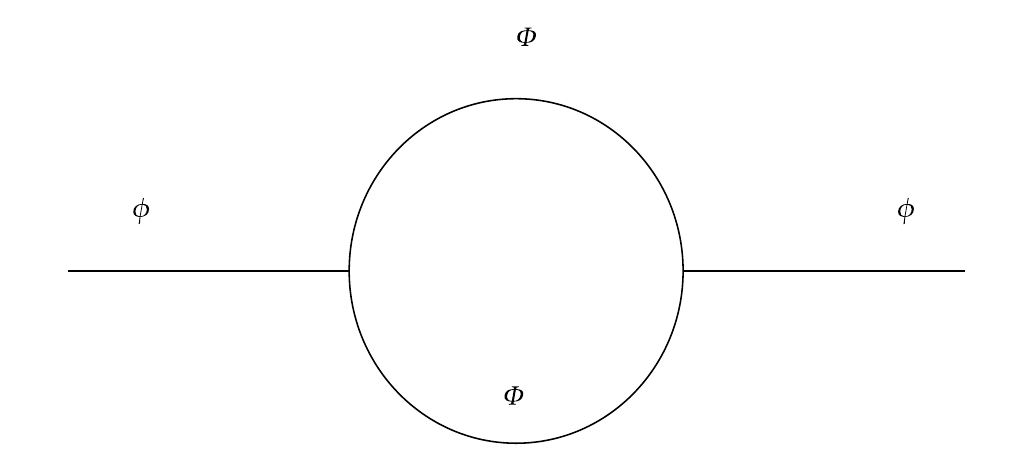}
    \caption{The self-energy Feynman diagram for $\phi\rightarrow\Phi\Phi$ process, which exist in the Kadanoff-Baym equation.}
    \label{SELFENERGY}
\end{figure}
Using this Feynman diagram, one can write down the self-energy as
\begin{equation}
    \Pi^>(x,y)=-Y(x)Y(y)G_\Phi^>(x,y)G_\Phi^>(x,y)
\end{equation}
The Wigner transformation of this self-energy is 
\begin{equation}\label{seKB}
\begin{split}
    \Pi^>(k,X)&=-\int d^4r  \int\frac{d^4p}{(2\pi)^4}  G_\Phi^>(p,X)
    \int\frac{d^4p'}{(2\pi)^4}  G_\Phi^>(p',X)
    \\
    &\quad\quad\quad\quad\quad\quad\quad\quad\quad\quad\quad\quad\quad\quad\times Y(X+\frac{r}{2})Y(X-\frac{r}{2})e^{i(k-p-p')\cdot r}
    \\
    &=-\int\frac{d^4p}{(2\pi)^4}\int\frac{d^4p'}{(2\pi)^4}Y(X-\frac{i}{2}\partial_k)Y(X+\frac{i}{2}\partial_k)\delta^4(k-p-p')
    \\
    &\quad\quad\quad\quad\quad\quad\quad\quad\quad\quad\quad\quad\quad\quad\quad\quad\quad\quad\quad\quad\times G_\Phi^>(p,X) G_\Phi^>(p',X)\ \ .
\end{split}
\end{equation}
Substituting this expression into Eq.(\ref{FPCKBE}) with the replacement  $k\rightarrow k\pm\frac{i}{2}d_z$, $z\rightarrow z\mp\frac{i}{2}d_{k_z}$ in Eq.\eqref{seKB}, we obtain the corresponding collision term
\begin{equation}\label{KBcol}
\begin{split}
    e^{-i\Diamond}\{\Pi^>,G^<\}=-\int\frac{d^4p}{(2\pi)^4}&\int\frac{d^4p'}{(2\pi)^4} (2\pi)^4Y(\hat{X}_{+})Y(\hat{X}_{-})\delta^4(k+\frac{i}{2}d_z-p-p')
    \\
    &\quad\quad\quad\quad\times G_\Phi^>(p,z-\frac{i}{2}\partial_{k_z}) G_\Phi^>(p',z-\frac{i}{2}\partial_{k_z})G_\phi^<(k,z)
    \\
    -\int\frac{d^4p}{(2\pi)^4}&\int\frac{d^4p'}{(2\pi)^4} (2\pi)^4Y(\hat{X}_{+})Y(\hat{X}_{-})\delta^4(k-\frac{i}{2}d_z-p-p')
    \\
    &\quad\quad\quad\quad\times G_\Phi^>(p,z+\frac{i}{2}\partial_{k_z}) G_\Phi^>(p',z+\frac{i}{2}\partial_{k_z})G_\phi^<(k,z)
\end{split}
\end{equation}
where $\hat{X}_\pm=X\pm\frac{i}{2}\partial_k$ that acts on the $\delta$ functions. 

Before proceeding further, we note that the derivatives appearing in Eq.\eqref{KBcol} contain two parts: the derivatives inside the Greens function, 2. the derivatives acting or inside the $\delta$-functions. 
The derivative inside the Green's functions can be simplified by first integrating the temporal component of the momentum, then expanding the term in powers of the derivative, which brings a factor of $\epsilon_{wall}$ as a perturbation parameter, just as in Eq.\eqref{eq:fided}
\begin{equation}
\begin{split}
    \int_0^\infty \frac{dp^0}{2\pi}\int_0^\infty \frac{dk^0}{2\pi}G_{\phi/\Phi}^>(p,z\pm\frac{i}{2}\partial_{k_z})G^<_\phi(k,z)&=\frac{f_{\phi/\Phi}(p,z\pm\frac{i}{2}\partial_{k_z})}{2E_p(z)}\frac{f_\phi(k,z)}{E_k(z)}
    \\
    &=\frac{f_{\phi/\Phi}(p,z)}{2E_p(z)}\frac{f_\phi(k,z)}{E_k(z)}+\mathcal{O}(\epsilon_{wall}).
\end{split}
\end{equation}
However, the derivatives acting on or inside the $\delta$-functions introduce a singularity that cannot be further simplified. Thus, in general, we should not directly omit the derivative related to the $\delta$-function by simply performing a power-series expansion, truncating at a given order, and assuming that the remaining terms are proportional to higher powers of $\epsilon_{wall}$. In this sense, Approximation-1 means using the traditional $\epsilon_{wall/coll}$ perturbative expansion of the Green's functions, truncating that expansion at LO,  and keeping all non-trivial background field contributions inside the $\delta$-function as a realization of momentum non-conservation. 

Mathematically speaking, we can Taylor expand $G^a(p,X-\frac{i}{2}\partial_k^b)G^b(k,X)$ , and the result is proportional to $\epsilon$ as the normal case~\cite{Cirigliano:2009yt}. On the contrary, we cannot omit any derivative related to the $\delta$-function since this expansion will include the summation of an infinite series of $\delta$-functions and their derivatives, and the convergence of such an expansion is questionable at best. Furthermore, under this approximation, it is reasonable that Approximation-2 makes sense because we only expand Green's function to LO/NLO in the constraint and kinetic KBE's; the {justification of the Approximation-2}  is given in the Appendix.~\ref{APPE}. In this sense, we formulate a non-perturbative treatment that focuses on grasping information about the momentum non-conservation through the resummation of $\epsilon_{wall}$ to all orders. As we find below, the friction pressure given by the BQFT method is a reflection of the the non-perturbative content of the KBE, it is not surprising that we do not see this effect using the naive perturbative expansion.

Now, let us proceed to simplify the Eq.\eqref{KBcol}. In light of the foregoing discussion, we need to keep all terms in $Y(\hat{X}_{+})Y(\hat{X}_{-})\delta^4(k+\frac{i}{2}d_z-p-p')$ but expand the Green's function to drop all the derivatives as Eq.\eqref{eq:fided} throught the Approximation-1.
To go further, we must deal with the derivative inside the delta function. If $G_\phi^b(k,z)$ is a non-trivial function of the $z$-coordinate, then four-momentum conservation will be broken, generating friction pressure coming from the collision term in the KBE. To extract the non-trivial information inside the $\delta$-functions, we can use the Wigner transformation acting on the average coordinate 
\begin{equation}\label{NWigner}
    G^\gtrless(k,l)=\int d^4 X e^{il\cdot X}G^\gtrless(k,X)\quad\quad G^\gtrless(k,X)=\int \frac{d^4l}{(2\pi)^4} e^{-il\cdot X} G^\gtrless(k,l).
\end{equation}

Here, we have assumed that $G^\gtrless(k,z)$ is only a function of $z$-coordinate. In this case, we only need to Wigner transform for $z$, which is 
\begin{equation}
G^\gtrless(k,z)=\int \frac{dl_z}{2\pi} e^{-il_z z} G^\gtrless(k,l_z) \ \ \ .
\end{equation}
Substituting it into Eq.(\ref{KBcol}) yields
\begin{equation}
\begin{split}
    e^{-i\Diamond}\{\Pi^>,G^<\}=-\int&\prod_{a=p,p'}\frac{d^4a}{(2\pi)^4}\int\frac{dl_z}{2\pi}(2\pi)^4\delta^3(\vec{k}_n-\vec{p}_n-\vec{p}'_n)Y(\hat{X}_{+})Y(\hat{X}_{-})[
    \\
    &\quad\times \delta(k_z+\frac{l_z}{2}-p_z-p'_z)]G_\Phi^>(p,z) G_\Phi^>(p',z)G_\phi^>(k',z)e^{-il_z z}
    \\
    -\int&\prod_{a=p,p'}\frac{d^4a}{(2\pi)^4}\int\frac{dl_z}{2\pi}(2\pi)^4\delta^3(\vec{k}_n-\vec{p}_n-\vec{p}'_n)Y(\hat{X}_{+})Y(\hat{X}_{-})[
    \\
    &\quad\times \delta(k_z-\frac{l_z}{2}-p_z-p'_z)]G_\Phi^>(p,z) G_\Phi^>(p',z)G_\phi^<(k,l_z)e^{-il_z z}
\end{split}
\end{equation}
where $\delta^3(\vec{k}_n-\vec{p}_n-\vec{p}'_n)\equiv\delta(k_0-p_0-p'_0)\delta^2(\vec{k}_\perp-\vec{p}_\perp-\vec{p}'_\perp)$. The momentum non-conservation effect appears via the $\delta$-functions containing $l_z$, as appearance of the latter clearly implies $k_z\not=p_z+p'_z$.
Performing the $l_z$ integration and using the $\delta$-functions leads to
\begin{equation}\label{eq.exp}
\begin{split}
    e^{-i\Diamond}\{\Pi^>,G^<\}=-&\int\frac{d^4p}{(2\pi)^4}\int\frac{d^4p'}{(2\pi)^4} (2\pi)^3\delta^3(\vec{k}_n-\vec{p}_n-\vec{p}'_n)Y(\hat{Z}_{+})Y(\hat{Z}_{-})
    \\
    &\quad\quad\quad\quad\quad\quad\quad\times [G_\Phi^>(p,z) G_\Phi^>(p',z)G_\phi^<(k,2\Delta p'_z)e^{-2i\Delta p_z z}]
    \\
    -&\int\frac{d^4p}{(2\pi)^4}\int\frac{d^4p'}{(2\pi)^4} (2\pi)^3\delta^3(\vec{k}_n-\vec{p}_n-\vec{p}'_n)Y(\hat{Z}_{+})Y(\hat{Z}_{-})
    \\
    &\quad\quad\quad\quad\quad\quad\quad\times [G_\Phi^>(p,z) G_\Phi^>(p',z)G_\phi^<(k,-2\Delta p'_z)e^{2i\Delta p_z z}]
\end{split}
\end{equation}
where $\hat{Z}_\pm=z\pm\frac{i}{2}\partial_{\Delta p_z}$ and $\Delta p_z=k_z-p_z-p'_z$.

We can now use the Approximation-2 to write down the form of the Green's function by Eq.(\ref{SCKBE}). The time component momentum $k_0,p_0, p'_0$ in the phase space integration can be separated into two regions $k_0>0$ or $k_0<0$. As shown in Eq.\eqref{SCKBE}, the two regions of the phase space correspond to in or out state in statistical physics, with corresponding dependencies on the distribution functions $f(k,z)$ or $1\pm f(k,z)$. The different choices of the integrated region correspond to different scattering processes. For example, $k_0,p_0,p'_0>0$ corresponds to the scattering process shown in the right panel of Fig.\ref{micro}. The total phase space integration  contains all scattering processes. Substituting Eq.\eqref{eq.exp} and the distribution function into Eq.(\ref{FPCKBE}) and setting the integration region for time component momentum as $k_0,p_0,p'_0>0$ , we have
\begin{equation}\label{OKBEF}
\begin{split}
    \mathcal{C}_1=-\int_0^\infty\frac{dk^0}{2\pi}k_z\int\frac{d^3\vec{p}}{(2\pi)^3 2E_p}\int\frac{d^3\vec{p}'}{(2\pi)^3 2E_{p'}}[1+f_\Phi(p,z)][1+f_\Phi(p',z)]Y(\hat{Z}_{+})Y(\hat{Z}_{-})
    \\
    \times [G_\phi^<(k,2\Delta p_z)e^{-2i\Delta p_z z}](2\pi)^3\delta^3(\vec{k}_n-\vec{p}_n-\vec{p}'_n) +(\Delta p_z\leftrightarrow-\Delta p_z)
\end{split}
\end{equation}
where $p_0$ and $p'_0$ have been integred out using the delta functions inside Eq.(\ref{SCKBE}). The Wigner transformation $G^<(k,l)$ can be given by the definition Eq.(\ref{NWigner}) as
\begin{equation}
    G^<_\phi(k,l_z)=\int dz' e^{il_z z'}(2\pi)\delta[k^2-m(z')^2](\Theta(k_0)f_\phi(k,z')+\Theta(-k_0)[1+\bar{f}(-k,z')])
\end{equation}
Substituting this expression for $G^<_\phi(k,l_z)$ back to  Eq.(\ref{OKBEF}), we obtain
\begin{equation}\label{eq:fricKB}
\begin{split}
    \mathcal{C}_1=-k_z&\int\frac{d^3\vec{p}}{(2\pi)^3 }\int\frac{d^3\vec{p}'}{(2\pi)^3}
    \frac{1+f_\Phi(p,z)}{2E_{p}}
    \frac{1+f_\Phi(p',z)}{2E_{p'}}
    \\
    &\quad\quad\times\int dz' Y(z')Y(2z-z')\frac{f_\phi(k,z')}{2E_k}e^{-2i\Delta p_z (z-z')}(2\pi)^3\delta^3(\vec{k}_n-\vec{p}_n-\vec{p}'_n)
    \\
    +(\Delta &p_z\leftrightarrow-\Delta p_z)
\end{split}
\end{equation} 
where we have used that $Y(\hat{Z}_{\pm})e^{-2i\Delta p_z (z-z')}=Y[z\pm(z-z')]e^{-2i\Delta p_z (z-z')}$. 

\noindent D. The fourth term $\mathcal{C}_2$ can be treated in the same way. However, since Green's function $G^h(k,z)$ in $\mathcal{C}_2$ is higher order compared with the $G^\gtrless(k,z)$ in $\mathcal{C}_1$, we conclude that the dominant collision terms that contain the momentum non-conservation effect arise in $\mathcal{C}_1$ and neglect the contribution from $\mathcal{C}_2$.



With the foregoing discussion in mind, now we can write down the modified non-local Boltzmann equation from the Kadanoff-Baym equation as
\begin{equation}\label{MBE}
\begin{split}
    \bigg[2k_z&\frac{\partial}{\partial z}-\frac{dm^2(z)}{dz}\frac{\partial}{\partial k_z} \bigg]\frac{f_\phi(k,z)}{E_k}
    \\
    &=-\int\frac{d^3\vec{p}}{(2\pi)^3 }\int\frac{d^3\vec{p}'}{(2\pi)^3}
    \frac{F(k,z)}{E_k}\frac{1+f_\Phi(p,z)}{2E_{p}}\frac{1+f_\Phi(p',z)}{2E_{p'}}
    \\
    &\quad\quad\quad\times(2\pi)^3\delta(E_k-E_p-E_{p'})\delta^2(\vec{k}_\perp-\vec{p}_\perp-\vec{p}'_\perp)+(\Delta p_z\leftrightarrow-\Delta p_z)
    \\
    &+{\rm Inverse Process}\ \ \ ,
\end{split}
\end{equation}
where we have defined the function $F(k,z)$ as
\begin{equation}\label{eq:inter}
    F(k,z)=\int dz'  f_\phi(k,z')Y(z')Y(2z-z')e^{-2i\Delta p_z (z-z')}\ \ \ .
\end{equation}
Eq.\eqref{MBE} is the modified Boltzmann equation, which includes the momentum non-conservation effect. 
If one ignores all non-local momentum structures and coordinate dependence of coupling, then one can prove that $F(k,z)\rightarrow Y^2f_{\phi}(k,z)(2\pi)\delta(\Delta p_z)$ and Eq.~(\ref{MBE}) will return to the ordinary Boltzmann equation. To demonstrate the non-trivial physics of Eq.\eqref{MBE}, we can compare it with Eq.\eqref{BE:COMPARE}. By observing these two equations, one can immediately find that if we replace $F(k,z)$ in Eq.\eqref{MBE} with $f_a(k,z)|M_{a\rightarrow bc}|^2/L$, one can obtain Eq.\eqref{BE:COMPARE}. To get a better understanding, we can write the amplitude in Eq.\eqref{BE:COMPARE} by using its definition at BQFT discussion above, leading to
\begin{eqnarray}\label{BQFT:AMP}
\begin{split}
    f_a(k,z)\frac{|M_{a\rightarrow bc}|^2}{L}&=f_a(k,z)\frac{|\int dz \chi_k^*(z)Y(z)\chi_p(z)\chi_{p'}(z)|^2}{L}
    \\
    &=f_a(k,z)\frac{\int dz\int dz'Y(z)Y(z')e^{-i(k_z-p_z-p_z')(z-z')}}{L}.
\end{split}
\end{eqnarray}
By comparing Eq.,\eqref{BQFT:AMP} and Eq.\eqref{eq:inter}, we can find that the only differences between those two equations are how we describe the amplitude. In Eq.\eqref{MBE}, the distribution function $f_\phi(k,z')$ is always linked with the the non-local momentum structure in Eq.\eqref{eq:inter}. Thus, it is a non-local integro-differential equation for the distribution function. On the contrary, Eq.\eqref{BE} is a local equation of the distribution function although the interaction in the BQFT method contains momentum non-conservation brought by the non-local background field. But we will prove in the next subsection that two equations give the same friction force under Approximation-3.



\subsection{The Friction Pressure from the Kadanoff-Baym Equation}
To obtain the friction pressure in background field QFT Eq.\eqref{fricGn} from Eq.\eqref{MBE}, one can follow the steps in  Section \ref{sec:BtoF}. Firstc, we need to multiply the $k_z$ and integrate the equation by $\int dz\int d^3\veck/(2\pi)^3$. This step would lead the RHS of the equation to become the friction pressure in the Eq.\eqref{eq:fricKB}. Then using Approximation-3, which reduces the coordinate dependence of the distribution function $f_a(k,z)\rightarrow f_a(k)$ and also using the classical approximation $1+f\approx 1$
\begin{equation}\label{FKBEC}
\begin{split}
    \frac{F^{fric}_{KB}}{A}&=-\int\frac{d^3\vec{k}}{(2\pi)^3 }\frac{f^{in}_\phi(k)}{2E_k}\int\frac{d^3\vec{p}}{(2\pi)^3 2E_p}\int\frac{d^3\vec{p}'}{(2\pi)^3 2E_{p'}}|A|^2(k_z-p_z-p'_z)
    \\
    &\quad\quad\quad\quad\quad\quad\quad\quad\quad\quad\quad\quad\quad\times(2\pi)^3\delta(E_k-E_p-E_{p'})\delta^2(\vec{k}_\perp-\vec{p}_\perp-\vec{p}'_\perp)
    \\
    &+{\rm Inverse Process}.
\end{split}
\end{equation} 
where $|A|^2$ is given by
\begin{equation}
\begin{split}
    |A|^2&=2\int dz\int dz' Y(z')Y(2z-z') e^{-2i\Delta p_z (z-z')}
    \\
    &=\int d(2z-z')\int dz' Y(z')Y(2z-z') e^{-i\Delta p_z (2z-z')}e^{-i\Delta p_z z'}
    \\
    &=\bigg|\int dz Y(z) e^{-i\Delta p_z z}\bigg|^2.
\end{split}
\end{equation}
Notice the overall factor $2$ comes from the friction pressure which is brought by the term $\Delta p_z\leftrightarrow-\Delta p_z$. The friction pressure from Eq.(\ref{FKBEC}) has the same form as the friction pressure Eq.\eqref{spliting} and Eq.(\ref{fricGn}), which is from the BQFT method. The amplitude in the BQFT method is given by Eq.(\ref{amplitude}) and in the WKB approximations, we have
\begin{equation}\label{Prove:equal}
\begin{split}
    |M|_{\phi\rightarrow\Phi^2}^2\approx\Bigg|\int dz Y(z) e^{i\Delta p_z z}\Bigg|^2=|A|^2.
\end{split}
\end{equation}
This result means the friction pressure for $\phi\rightarrow\Phi^2$ from the BQFT method can be automatically generated by the collision term in the Kadanoff-Baym equation. As a consequence, we obtain the friction pressure in the Eq.\eqref{FKBEC}, which has been proved to be the same friction pressure in background field QFT, in the Eq.\eqref{Prove:equal}.


To end this section, let us summarize the logic and the corresponding approximation we used to get the friction force and modified Boltzmann equation from the Kadanoff-Baym equation. In principle, one can solve the complete Kadanoff-Baym equations Eq.(\ref{KBE}) to find the answer for bubble wall-velocity. But it is not necessary to solve the complete Kadanoff-Baym equation to get the non-local effect. We apply the WKB/$\epsilon_{wall}$ expansion to the Green's function, which lead us to Eq.(\ref{MBE}). This equation contain all the momentum non-conservation effect of the bubble wall and particle interactions. 
Finally, if we are only interested in the microscopic friction force, we can do the thermal equilibrium approximations on the distribution function and find the result.

\section{The $2\rightarrow 2$ Friction Force and the Bubble Wall Velocity}\label{22preV}
As an application of Eq.(\ref{fricGn}) or Eq.(\ref{FKBEC}), rather than only consider the $1\rightarrow n$ contribution like the splitting process, we can use it to study a friction force arising from the $2\rightarrow 2$ process, which has not been applied in previous analyses.
From Eq.(\ref{fricGn}) or  Eq.(\ref{FKBEC}), we can written down the friction pressure of $2\rightarrow 2$ process as
\begin{equation}\label{22phic}
\begin{split}
    \frac{F_{2\rightarrow 2}^{fric}}{A}&=\int\frac{d^3\vec{p}}{(2\pi)^3}\frac{f(p)}{2E_p}\int\frac{d^3\vec{p}'}{(2\pi)^3}\frac{f(p')}{2E_p'}\int\frac{d^3\vec{k}}{(2\pi)^3}\frac{1}{2E_k}\int\frac{d^3\vec{k}'}{(2\pi)^3}\frac{1}{2E_k'}|M|^2_{2\rightarrow2}(p_z+p_z'-k_z-k_z')
    \\
    &\quad\quad\quad\quad\quad\quad\quad\quad\quad\quad\quad\quad\times(2\pi^3)\delta(E_p+E_p'-E_k-E_k')\delta^{(2)}(\vec{p}_\perp+\vec{p}'_\perp-\vec{k}_\perp-\vec{k}'_\perp)
    \\
    &+{\rm Inverse\ Process}.
\end{split}
\end{equation}
where $M_{2\rightarrow 2}$ is the amplitude for $2\rightarrow 2$ process and should be study case by case. Integrating out all the delta functions, we can get
\begin{equation}\label{pre22}
\begin{split}
    \frac{F_{2\rightarrow 2}^{fric}}{A}&=\int\frac{d^3\vec{p}}{(2\pi)^3}\frac{f(p)}{2E_p}\int\frac{d^3\vec{p}'}{(2\pi)^3}\frac{f(p')}{2E_p'}\int\frac{d^3\vec{k}}{(2\pi)^3}\frac{1}{2E_k2k_z'}|M|^2_{2\rightarrow2}(p_z+p_z'-k_z-k_z')
    \\
    &+{\rm Inverse\ Process}.
\end{split}
\end{equation}


To start our discussion, let us consider a two-scalar field portal interaction as 
\begin{equation}\label{laportal}
    \mathcal{L}_{int}=\lambda\phi^2\Phi^2
\end{equation}
where $\phi$ is a light scalar field with mass $m_\phi$ and $\Phi$ is a heavy scalar field with mass $m_\Phi$ where $m_\phi< m_\Phi$. This model is prevalent in the BSM model, also called as the Higgs portal model, which generates the FOEWPT~\cite{OConnell:2006rsp, OConnell:2006rsp,Barger:2008jx, Profumo:2014opa, Chiang:2017nmu, Wang:2023zys}. For example, the phase transition in the complex scalar extension of the standard model~\cite{Chiang:2017nmu}, where $s$ indicates $\Phi$ and $H$ represents $\phi$. For the Feynman diagram, which is presented in the left panel of Fig.\ref{SELFENERGY2}, the amplitude can be written by
\begin{equation}\label{amp22phi}
    M_{2\rightarrow2}=\int dz \chi^*_k(z)\chi^*_{k'}(z)V(z)\chi(k)\chi_p(z)\chi_{p'}(z).
\end{equation}
where the vertex function is given by $V(z)=-i\lambda$.
\begin{figure}[htbp] 
    \centering
    \includegraphics[width=0.3\textwidth]{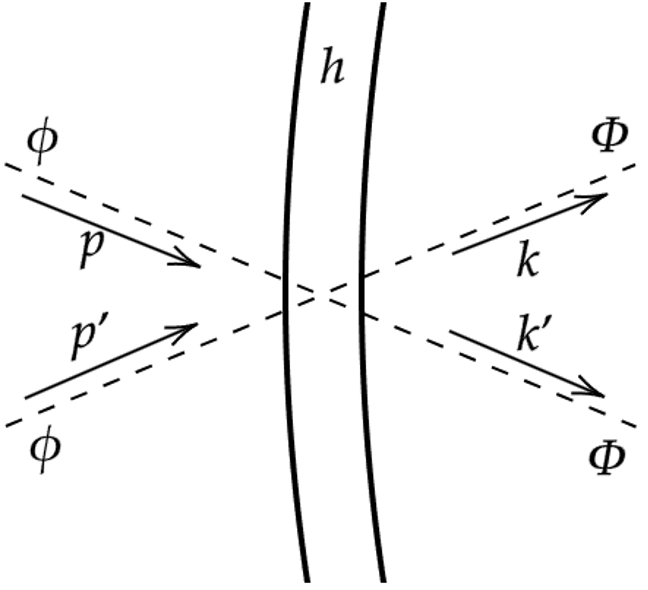} 
    \includegraphics[width=0.5\textwidth]{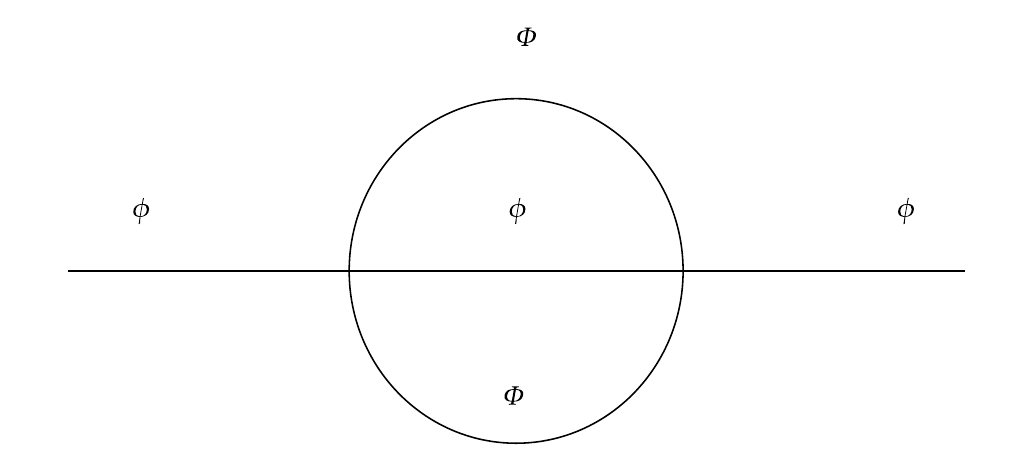}
    \caption{Left panel: The Feynman diagram for scattering process which generates the $2\rightarrow 2$ friction pressure. Right panel: The corresponding self-energy Feynman diagram for this process.}
    \label{SELFENERGY2}
\end{figure}
We will use the same method in \cite{Bodeker:2017cim} to compute this amplitude. Firstly, as mentioned in Section.\ref{IncS}, the non-trivial contribution of the friction pressure is only generated at large wall velocity where the local thermal equilibrium is invalid. Thus, we want to consider that bubble velocity is close to the speed of light $\gamma_w\gg 1$. In this case, the momentum in $z$ direction will satisfied $p_z\sim E_p\gg|\vec{p}_\perp|\sim m_\Phi\ {\rm or}\ m_\phi$. Using this information, we can expand the mode function in amplitude as
\begin{equation}
\begin{split}
    \chi^*_k(z)\chi^*_{k'}(z)\chi_p(z)\chi_{p'}(z)=\exp\bigg[\frac{i}{2}\int_0^zdz'\bigg(\frac{\vec{p}_\perp^2+m_1^2}{E_p}&+\frac{\vec{p}_\perp'^2+m_1^2}{E_{p}'}
    \\
    &-\frac{\vec{k}_\perp^2+m_2^2}{E_k}-\frac{\vec{k}_\perp'^2+m_2^2}{E_k'}\bigg)\bigg]
\end{split}
\end{equation}
To simplify this problem, let us use approximations for incoming particles. Suppose the momentum for the incoming particle is on the z-axis 
and the width of the wall is thin compared to the interaction scale located at $z=0$ then, we can rewrite the amplitude as
\begin{equation}
    M_{2\rightarrow2}=-i\lambda\lim_{L\rightarrow0}\bigg(\int_{-\infty}^{-\frac{L}{2}} dz+\int_{\frac{L}{2}}^{\infty} dz\bigg)\chi^*_k(z)\chi^*_{k'}(z)\chi(k)\chi_p(z)\chi_{p'}(z).
\end{equation}
One should notice that there are three processes which would contribute to the friction pressure, $\phi^2\rightarrow\Phi^2$, the inverse process $\Phi^2\rightarrow\phi^2$ and the mixied scattering process $\phi\Phi\rightarrow\phi\Phi$. Using the form of the mode function, we can find the amplitude two those two processes as
\begin{equation}\label{MMphi}
\begin{aligned}
    &M_{\phi^2\rightarrow\Phi^2}=\frac{2i\lambda}{E_p+E_p'}\bigg[\bigg(\frac{{m^h_\phi}^2}{E_pE_p'}-\frac{\vec{k}_\perp^2+{m^h_\Phi}^2}{E_kE_k'}\bigg)^{-1}-\bigg(\frac{{m^s_\phi}^2}{E_pE_p'}-\frac{\vec{k}_\perp^2+{m^s_\Phi}^2}{E_kE_k'}\bigg)^{-1}\bigg],
    \\
    &M_{\Phi^2\rightarrow\phi^2}=\frac{2i\lambda}{E_p+E_p'}\bigg[\bigg(\frac{{m^h_\Phi}^2}{E_pE_p'}-\frac{\vec{k}_\perp^2+{m^h_\phi}^2}{E_kE_k'}\bigg)^{-1}-\bigg(\frac{{m^s_\Phi}^2}{E_pE_p'}-\frac{\vec{k}_\perp^2+{m^s_\phi}^2}{E_kE_k'}\bigg)^{-1}\bigg]. 
    \\
    &M_{\phi\Phi\rightarrow\phi\Phi}=2i\lambda\bigg[\bigg(\frac{{m_\phi^h}^2E_p'+{m^h_\Phi}^2E_p}{E_pE_p'}-\frac{\vec{k}_\perp^2(E_p+E_p')+{m^h_\phi}^2E_k'+{m^h_\Phi}^2E_k}{E_kE_k'}\bigg)^{-1}
    \\
    &\quad\quad\quad\quad\quad\quad-\bigg(\frac{{m_\phi^s}^2E_p'+{m^s_\Phi}^2E_p}{E_pE_p'}-\frac{\vec{k}_\perp^2(E_p+E_p')+{m^s_\phi}^2E_k'+{m^s_\Phi}^2E_k}{E_kE_k'}\bigg)^{-1}\bigg]. 
\end{aligned}
\end{equation}
 Before proceeding further with the calculation, let us briefly summarize the approximation which we would like to use in the following computation. The amplitude for $\phi^2\rightarrow\Phi^2$ can be simplified by assuming that $m_\phi^s$ is much smaller than other scales(corresponding to zero mass in symmetric phase), which justifies dropping the terms proportional to $m_\phi^s$ in Eq.\eqref{MMphi}. In addition, we also assume that $m_\Phi>m_\phi$ which implies $\veck^2_\perp+{m_\Phi^h}^2>{m_\phi^h}^2$. So, $\veck^2_\perp+{m_\Phi^h}^2$ becomes the dominant term in the amplitude $M_{\phi^2\rightarrow\Phi^2}$. The inverse process is more complicated since $\veck_\perp^2+m_\phi^2$ is not necessarily smaller than $m_\Phi^2$. So, to simplify the computation, let us separate the integral region of the phase space as $\veck_\perp^2+m_\phi^2<m_\Phi^2$ and $\veck_\perp^2+m_\phi^2>m_\Phi^2$ . In each region, we keep only the dominant contribution.
After some cumbersome integration and algebra simplification, which can be found in Appendix.\ref{App:Com}, we find the dominant terms in friction pressure in $\phi^2\leftrightarrow\Phi^2$ are given by
\begin{equation}\label{fric22c3}
\begin{split}
    \frac{F_{\phi^2\leftrightarrow\Phi^2}^{fric}}{A}&\approx    I_1(\gamma_w,T)\times I_2
\end{split}
\end{equation}
where the expression $I_1$ and $I_2$ is given by
\begin{equation}\label{eq:III}
\begin{split}
    &I_1(\gamma,T)=\int \frac{d^3\vec{p}}{(2\pi)^3}\int \frac{d^3\vec{p}'}{(2\pi)^3}\int \frac{dk_z}{2\pi} \frac{p_z p_z'f_\Phi(p)f_\Phi(p')}{(p_z+p_z')(k_z^2+{m_\phi^h}^2)(p_z+p_z'-\sqrt{k_z^2+{m^h_\phi}^2})^2}
    \\
    &I_2=\int\frac{d^2\vec{k}_\perp}{(2\pi)^2}\frac{\lambda^2\Delta m_\Phi^4(\vec{k}_\perp^2+ m_\phi^2)}{16{m_\Phi^h}^4{m_\Phi^s}^4}\Theta({m_\Phi^h}^2-\vec{k}^2_\perp-{m_\phi^h}^2)
\end{split}
\end{equation}
the $f(p)\approx \exp[-\gamma(E_p-v_w p_z)/T]$ is the distribution function of incoming particles. To compute the concrete numerical result, let us simplify this integral by ignoring the $\vec{p}_\perp$ in the $E_p$ except the energy inside the distribution function. Then, the value of the friction pressure can be obtained from numerical integration and is shown as the Green solid line in the Fig.\ref{pcompare}.
\begin{figure}[t] 
    \centering
    \includegraphics[width=0.45\textwidth]{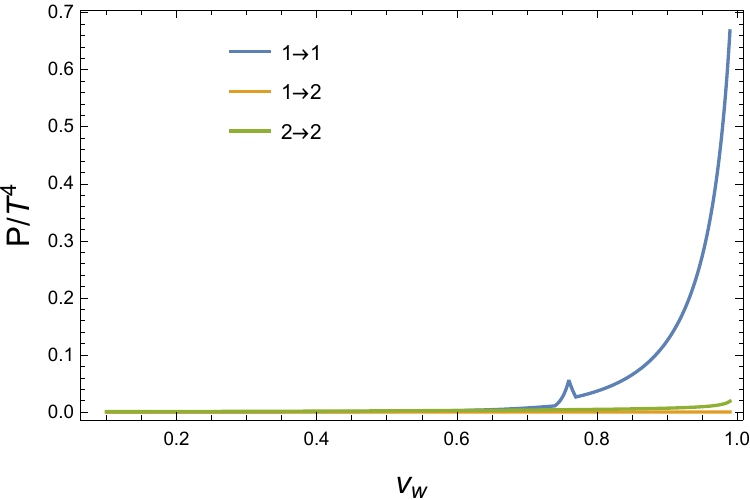}
    \includegraphics[width=0.45\textwidth]{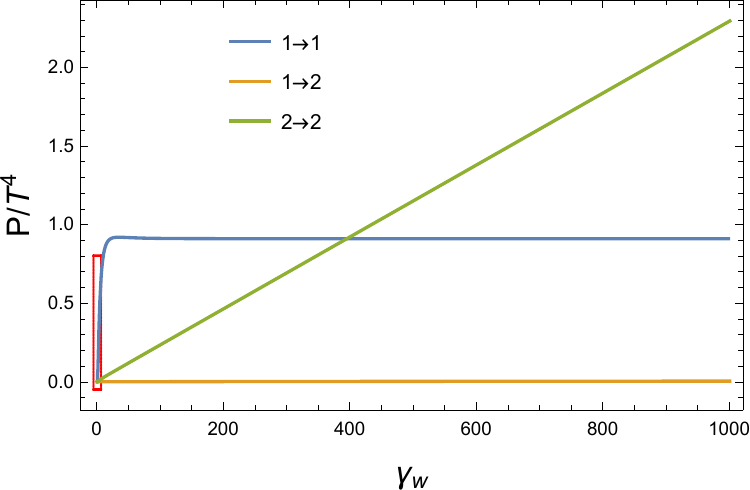} 
    \caption{The comparison between three friction pressures, where we set $\lambda\sim 1$ and $m_\phi^h=m_\Phi^s\approx T$ and $m_\Phi^h=10T$. The $1\rightarrow1$ classical friction pressure from the mass variation is labeled by the Blue solid line. The $ 1\rightarrow 2$ friction pressure from the splitting process is labeled by the Yellow solid line. The $2\rightarrow2$ friction force is labeled by the Green solid line. The left panel is a plot for $0<v_w<0.995$ to demonstrate the low $v_w$ region which is an enlargement of the red dotted box on the right panel; The right panel is the same plot but with more larger scale of $\gamma_w$($1\leq\gamma_w\leq 1000$).}
    \label{pcompare}
\end{figure} 
To show the importance of the $2\rightarrow 2$ friction pressure. We also present a comparison between three kinds of friction pressure in prue scalar theory in Fig.\ref{pcompare}: \begin{itemize}
    \item The classical friction force $P_{\Phi_s\rightarrow\Phi_h}$ provided by mass variation of $1\rightarrow 1$ process. It can be computed by the second term in the Eq.\eqref{Fric} under the relativistic ballistic approximation, and is nearly a constant with large wall velocity~\cite{Bodeker:2009qy}. 
    \item The friction force from the $\phi\rightarrow\Phi\Phi$ splitting process also exists in this pure scalar theory. The corresponding Lagrangian (Eq.\eqref{lasp}) can be obtained by replacing one field in Eq.\eqref{laportal} by the vacuum expectation background field(bubble). The corresponding friction pressure $P_{\phi_s\rightarrow\Phi^2_h}$ can be found in~\cite{Baldes:2022oev, Ai:2023suz}. Hence, we have
    \begin{equation}
        P_{\phi_s\rightarrow\Phi^2_h}\approx10^{-3}\lambda\log(1+0.26\frac{\gamma_w T}{L_w m_\Phi^2})P_{\Phi_s\rightarrow \Phi_h}(\gamma_w\rightarrow\infty)
    \end{equation}
    where $L_w$ is the width of the bubble, in this computation we choose it as $L_w\sim0.1$~\footnote{The value of $L_w$ is not very important for large range, because its effect is suppressed by the logarithm.}.
    \item The friction force brought by the $\Phi_s^2\rightarrow\phi_h^2$ scattering process in pure scalar theory(labeled by the Green solid line).
\end{itemize}
One should notice, we only consider the detonation fluid contribution in $P_{\Phi_s\rightarrow\Phi_h}$, which is the only possible configuration for fluid at large $v_w$. 
For the small $v_w$ region as shown in the left panel of Fig.\ref{pcompare}, other fluid configurations, for example, the deflagration, may become dominant.
If we count the effect of deflagration, we would find a larger friction pressure $P_{\Phi_s\rightarrow\Phi_h}$, for $v_w<v_J$ and as shown in~\cite{Ai:2024btx}, the transition from the deflagration into detonation fluid at $v_J$, will create a peak at $v_J$ just like in the first panel of Fig.\ref{frictionFvsM}.
However, to obtain the friction pressure from the deflagration configuration, we have to first assume a specific phase transition in the BSM model, then solve a group of complex fluid EOM, which is beyond the scope of this paper. So, we will leave this work for future study.

From Fig.\ref{pcompare} we know that the $2\rightarrow2$ friction pressure is dominated at the high $\gamma_w$ regions.
This result indicates that if the first-order electroweak phase transition contains two scalars with significant mass differences, the bubble will never run away, even if those scalars are not coupled with the transverse gauge field. The above conclusion is different from the understanding given by previous soft emission studies, which will lead to a friction pressure $F^{fric}_{soft}\propto\gamma_w$ only when the gauge field is present~\cite{,Bodeker:2017cim}. It is also not similar to the light-to-heavy scalar splitting which yield a $F^{fric}_{LH-S}\propto\log(1+a\gamma_w)$~\cite{Ai:2023suz}. So, it is a new type of friction force that will eliminate the run-away bubble.

\section{Conclusion and Outlook}\label{con}
The bubble wall terminal velocity is a key parameter characterizing the dynamics of a first order cosmological phase transition. It is particularly relevant to the resulting gravitational wave spectrum and, in the case of a first order electroweak phase transition, the viability of electroweak baryogenesis.
To date, two methods have been widely employed to determine it: the fluid and microscopic methods. However, these methods are inconsistent, owing to different assumptions about (non-)conservation of momentum in microscopic interactions at the bubble walls.
In previous literature, there does not appear to be a framework that self-consistently incorporates the key effects emphasized in each method. This situation impairs the theoretical robustness of GW and EWBG computations.

In this work, we have developed such a self-consistent framework by introducing the BQFT method in the Boltzmann equation, for the first time. Considering the background field effect, we find that the collision terms in the Boltzmann equation break momentum conservation in particle interactions in a direction normal to the bubble wall, thereby generating the friction force obtained in the microscopic method. Since the BQFT Boltzmann equation also includes the dynamics of the background field, it incorporates the physics emphasized in the fluid method. Thus, we obtain a self-consistent framework for the dynamics relevant to both methods.

We have also shown how one may obtain the BQFT from the more general Kadanoff-Baym equations under certain well-motivated approximations. 
Doing so requires performing a careful delineation of the impact of impact of derivative operators acting on or within Greens functions and distributions. The former typically lead to a systematic expansion the parameter $\epsilon_{wall}$\cite{Cirigliano:2009yt,Cirigliano:2011di}, whereas one must sum the latter to all orders in order to obtain the momentum non-conservation effect. The resulting \lq\lq non-local\rq\rq\, Kadanoff-Baym kinetic equation reduces to the same form as the BQFT Boltzmann equation. Then, we demonstrated that the microscopic friction force, like splitting friction, can  be automatically generated by the \lq\lq non-local\rq\rq\,  Kadanoff-Baym kinetic equation.


Finally, as an application, we illustrated the friction force coming from the $2\rightarrow 2$ process (scattering and annihilation) in pure scalar theory for the first time. These $2\rightarrow 2$ processes generate a new friction pressure proportional to $\gamma_w$, even without inclusion of a gauge field as discussed in earlier work. This result means that the bubble in the phase transition that contains two scalar fields may never run away, even in the absence of gauge interactions.

Looking to the future, it remains a challenge to solve Eq.(\ref{MBE}) or, even more ambitiously, the full Kadanoff-Baym equations. In principle, doing so can generate the friction force, which forms the lower panel of Fig.\ref{frictionFvsM}. For Eq.~(\ref{MBE}), the methods utilized in Refs.~\cite{Laurent:2022jrs,DeCurtis:2022llw,DeCurtis:2023hil} may prove tractable. 
Solving this equation will also tell us whether the non-local effect would affect the energy fraction, which is another critical ingredient in GW computations.
A further interesting application would be to dark matter generated during phase transitions. While many papers~\cite{Azatov:2021ifm,Baker:2019ndr,Chway:2019kft,Jiang:2023nkj,Ai:2024ikj} have analyzed this possibility using the microscopic method, it would be more convincing the start from the non-local Kadanoff-Baym kinetic equation or BQFT Boltzmann equation, thereby making the treatment of the problem analogous to that of conventional dark matter freeze-out computations. 


\noindent 
\underline{\it Note added:} \\ 
While we were finalizing this paper, we learned that another group (Wen-Yuan Ai, Matthias Carosi, Björn Garbrecht, Carlos Tamarit and Miguel Vanvlasselaer) has been working on similar topics~\cite{Ai:2025bjw}. Their paper was posted jointly with the initial posting of our work.

\vskip 0.2in

\noindent {\bf{Acknowledgements}}
We thank Wenyuan Ai, Bjorn Garbrecht, Zhaofeng Kang, Benoit Laurent, Yucheng Qiu, Tuomas V. I. Tenkanen, Yanda Wu, and Yeling Zhou for useful discussions during this work. This work was supported in part by the National Natural Science Foundation of China under grant no. 12375094 and W2441004.
\appendix
\section{The Matching Condition for Fluid Method}\label{matching}
If we do not consider the gravitational effect, then the total energy-momentum tensor for the scalar fluid system must be conserved
\begin{equation}\label{EMC}
    \partial_\mu T^{\mu\nu}=0.
\end{equation}
The total energy-momentum tensor for the scalar fluid system is given by
\begin{equation}\label{EMTPHI}
    T^{\mu\nu}=T^{\mu\nu}_\phi+T^{\mu\nu}_f\equiv\partial^\mu\phi\partial^\nu\phi-g^{\mu\nu}\left[\frac{1}{2}(\partial\phi)^2-V(\phi)\right]+T_f^{\mu\nu},
\end{equation}
where $T_f^{\mu\nu}$ is the energy-momentum for fluid system. If the fluid is ideal fluid the $T_f^{\mu\nu}$ can be write by
\begin{equation}
    T_f^{\mu\nu}=\omega_f u^\mu u^\nu-g^{\mu\nu} p_f.
\end{equation}
Using the above expression, we can rewrite the total energy-momentum tensor as
\begin{equation}
    T^{\mu\nu}=\partial^\mu\phi\partial^\nu\phi+\omega_f u^\mu u^\nu-g^{\mu\nu}\left[\frac{1}{2}(\partial\phi)^2-V(\phi)+p_f\right].
\end{equation}
In the planer approximation, we only need to consider Eq.(\ref{EMC}) when $\nu=t,z$. Considering the stable bubble configuration representing the bubble reaching the terminal velocity, we will find $\partial_z T^{z z}=\partial_z T^{z 0}=0$. Then, integrating this equation by $\int dz$, we will find that
\begin{equation}
    \Delta(\omega_f \gamma^2 v)=0\quad\quad\Delta(\omega_f\gamma^2v^2-p_f)=-\Delta V
\end{equation}
This equation matches the condition that the scalar fluid system must be satisfied at terminal velocity. If we define $p=p_f-V$ and $\omega=\omega_f=Tdp/dT=Tdp_f/dT$, then the above equation can be written into a more common form
\begin{equation}\label{mcg}
    \Delta(\omega \gamma^2 v)=0\quad\quad\Delta(\omega\gamma^2v^2-p)=0
\end{equation}

\section{The Boltzmann Equation}\label{bleq}
Suppose the particle moves at the bubble background field, with velocity $\vec{v}$. In this case, the coordinate of the particle will be changed as $\vec{x}'=\vec{x}+\vec{v}dt$, and the momentum of the particle will be changed as $\vec{p}'=\vec{p}+\vec{F} dt$. Following that information, we can construct the Boltzmann equation from the kinetic theory. The particle number variation can be written by
\begin{equation}
\begin{split}
    dN(t+dt)-dN(t)&=f(p',x',t+dt)d^3\vec{p}'d^3\vec{x}'-f(p,x,t)d^3\vec{p}d^3\vec{x}
    \\
    &=[\partial_t f(x,p,t)+v_i \partial_i f(x,p,t)+\partial_{p^i}(F^{i} f)]d^3\vec{p}d^3\vec{x}dt
\end{split}  
\end{equation}
The microscopic event can write the source for those number variations as
\begin{equation}
\begin{split}
    dN(t+dt)-dN(t)=-\prod_{i}N_i\frac{P}{T}dt
\end{split}  
\end{equation}
where $P$ is the probability of a microscopic event that changes the particle number and $N_i$ is the particle numbers that attend this event. So, the complete form of the kinetic equation for particle $a$ is given by
\begin{equation}
\begin{split}
    [\frac{p^\mu}{E_p} \partial_\mu f_a(x,p,t)+\partial_{p^i}(F^i f_a)]d^3\vec{p}d^3\vec{x}dt=-f_a(x,p,t)d^3\vec{p}d^3\vec{x}\prod_{i\neq a}N_i\frac{P}{T}dt
\end{split}
\end{equation}
where we have written the velocity as $v_i=p_i/E$, and the times are normalized by $T$. The probability of the microscopic event can be written by the initial and final states as
\begin{equation}
    dP=\frac{\langle f|i\mathcal{T}|i\rangle\langle i|i\mathcal{T}|f\rangle}{\langle i|i\rangle \langle f|f\rangle}\prod_{f}\frac{Vd^3\vec{p}_f}{(2\pi)^3}
\end{equation}
where $V$ is the volume of the space, the normalized of the initial and final states are written by
\begin{equation}
\begin{split}
    \langle i/f|i/f\rangle=\prod_{i/f}2E_{i/f}(2\pi^3)\delta^{(3)}(0)=\prod_{i/f}2E_{i/f}V
\end{split}
\end{equation}
where $V=L^3$ is the total space volume. Combined with all the information, we can write the Boltzmann equation as
\begin{equation}
\begin{split}
    \frac{p^\mu}{E_p} \partial_\mu f_a(x,p,t)+\partial_{p^i}[F^i f_a(x,p,t)]=\mathcal{C}[f_a],
\end{split}
\end{equation}
where the collision term can be written by
\begin{equation}\label{col}
    \mathcal{C}[f_a]=-f_a(x,p,t)\prod_{i\neq a}N_i\int\frac{\langle f|i\mathcal{T}|i\rangle\langle i|i\mathcal{T}|f\rangle}{T\prod_i (2E_i V)}\prod_{f}\frac{d^3\vec{p}_f [1\pm f_f(x,p_f,t)]}{2E_f(2\pi)^3}+{\rm Inverse\ Process},
\end{equation}
recognize the above equation; we can combine all particle numbers $N_i$ with the space volume $V$ into particle number density $n_i$ as
\begin{equation}\label{col}
    \mathcal{C}[f_a]=-f_a(x,p,t)\prod_{i\neq a}n_i\int\frac{\langle f|i\mathcal{T}|i\rangle\langle i|i\mathcal{T}|f\rangle}{T\prod_i (2E_i )V}\prod_{f}\frac{d^3\vec{p}_f[1\pm f_f(x,p_f,t)]}{2E_f(2\pi)^3}+{\rm Inverse\ Process},
\end{equation}
the remaining $V$ belongs to particle $a$. The above equation is the most general form of the collision terms. In ordinary QFT, this term will be the normal collision terms
\begin{equation}\label{ncol}
\begin{split}
    \mathcal{C}[f_a]=&-f_a(x,p,t)\prod_{i\neq a}\int\frac{d^3\vec{p}_i f_i(x,p_i,t)}{(2\pi)^3 2E_i}\prod_{f}\int\frac{d^3\vec{p}_f[1\pm f_f(x,p_f,t)]}{2E_f(2\pi)^3}|\mathcal{M}|_{i\rightarrow f}^2\delta^4(\sum_ip_i-\sum_fp_f)
    \\
    &+{\rm Inverse\ Process}.
\end{split}
\end{equation}

\section{The Quantization of the Field}\label{quantization}
The equation of the motion of the field in the background field is given by Eq.(\ref{EoMs}). Now, considering the scalar case with solution form as Eq.(\ref{solutionform}), we will get Eq.(\ref{SLE}). This is a standard Strum-Liouville equation with eigenvalue $\lambda$
\begin{equation}\label{SSLE}
    \frac{d}{dz}\left[k(z)\frac{df_\lambda(x)}{dz}\right]-q(z)f_\lambda(z)+\lambda^2 \rho(z) f_\lambda(z)=0
\end{equation}
where $k(z)=\rho(z)=1$ and $q(z)=m^2(z)$. Since $k(z)=\rho(z),q(z)>0$. If the solutions are real functions, then the eigenfunctions of this Sturm-Liouville equation will form an orthogonal complete set that satisfied
\begin{equation}
\begin{split}
    &\int dz f_\lambda(z) f_{\lambda'}(z)=2\pi\delta(\lambda-\lambda')
    \\
    &F(z)=\sum_\lambda a_\lambda f_\lambda(x)\rightarrow\int \frac{d\lambda}{2\pi} a_\lambda f_\lambda(z)
\end{split}
\end{equation}
the second equation means any function can be expanded by this complete set. For the first equation, we have to notice that for different eigenfunctions with the same eigenvalue, we force the $\delta$ function to equal zero. Because $f_\lambda$ is an orthogonal complete set, So, we can write down that
\begin{equation}
    \delta(x-a)=\int \frac{d\lambda}{2\pi} a_\lambda f_\lambda(x)
\end{equation}
to extract the information of $a_\lambda$, we can multiply the ${f_\lambda'}(x)$ in both side of this equation can integral over $x$. The result is given by
\begin{equation}
    f_{\lambda'}(a)=a_{\lambda'}
\end{equation}
Then, we will get the last property of this orthogonal complete set
\begin{equation}
    \delta(x-x')=\int \frac{d\lambda}{2\pi} f_\lambda(x)f_{\lambda'}(x')
\end{equation}
Normally, the solution $\chi_\lambda(z)$ is a complex function and can be written as
\begin{equation}
    \chi_\lambda(z)=\frac{x_\lambda(z)+iy_\lambda(z)}{\sqrt{2}}
\end{equation}
where $x_\lambda(z)$ and $y_\lambda(z)$ are both real function. And since all coefficients in Eq.(\ref{SSLE}) are real, then we know $x_\lambda(z)$ and $y_\lambda(z)$ both satisfied Eq.(\ref{SSLE}) and both can be treated as an orthogonal complete set for real functions. Owning this information, then let us prove the orthogonal relation of $\chi_\lambda$. We have
\begin{equation}\label{compofchi}
\begin{split}
    \int dz \chi_\lambda^*(z)\chi_{\lambda'}(z)&=\frac{1}{2}\int dz [x_\lambda(z)-iy_\lambda(z)][x_{\lambda'}(z)+iy_{\lambda'}(z)]
    \\
    &=\frac{1}{2}\int dz[x_\lambda(z)x_{\lambda‘}(z)+y_\lambda(z)y_{\lambda‘}(z)+ix_\lambda(z)y_{\lambda‘}(z)-iy_\lambda(z)x_{\lambda‘}(z)]
    \\
    &=\delta(\lambda-\lambda')+\frac{i}{2}\int dz[x_\lambda(z)y_{\lambda‘}(z)-y_\lambda(z)x_{\lambda‘}(z)]
\end{split}
\end{equation}
we must notice that both $x_\lambda$ and $y_\lambda$ are orthogonal complete set. So, we can expand $y_\lambda$ as $x_\lambda$ then we have
\begin{equation}
    y_{\lambda}(z)=\int \frac{d\alpha}{2\pi} g_{\lambda\alpha}x_\alpha(z)
\end{equation}
the orthogonal relation of $y_\lambda$ tell us that
\begin{equation}
\begin{split}
    2\pi\delta(\lambda-\lambda')&=\int dz y_\lambda(z) y_{\lambda'}(z)
    \\
    &=\int dz\int \frac{d\alpha}{2\pi}\int \frac{d\beta}{2\pi} g_{\lambda\alpha}g_{\lambda'\beta} x_\alpha(z)x_\beta(z)
    \\
    &=\int \frac{d\alpha}{2\pi} g_{\lambda\alpha}g_{\lambda'\alpha}
\end{split}
\end{equation}
where in the last line, we have used the orthogonal relation of $x_\lambda$. So, we have $g_{\lambda\alpha}\propto\delta(\lambda-\alpha)$. It may be a trick to understand this conclusion, notice that different eigenfunctions may have the same eigenvalue. So, the $\delta(\lambda-\alpha)$ means that they are the same eigenfunctions. For different eigenfunctions with the same eigenvalue, then $\delta(\lambda-\alpha)=0$. Substitute this result into Eq.(\ref{compofchi}), then the last term will cancel each other and we have
\begin{equation}
    \int dz \chi^*_\lambda(z)\chi_{\lambda'}(z)=2\pi\delta(\lambda-\lambda')
\end{equation}
Based on the same method we can also proof that 
\begin{equation}
    \int \frac{d\lambda}{2\pi} \chi_\lambda^*(x)\chi_\lambda(x')=\delta(x-x')
\end{equation}

Now, we can expand the field operator by those eigenfunctions as
\begin{equation}
\begin{split}
    \hat{\phi}(x)&=\int\frac{d\lambda d^2\vec{k}_\perp}{(2\pi)^3}\frac{1}{\sqrt{2\omega}}\left[a_{\lambda,k}\chi_\lambda(z)e^{-ik_n\cdot x_n}+a_{\lambda,k}^\dagger\chi_\lambda^*(z)e^{ik_n\cdot x_n}\right]_{\omega^2=\lambda^2+\vec{k}^2_\perp}
    \\
    &\approx\int\frac{d^3\vec{k}}{(2\pi)^3}\frac{1}{\sqrt{2\omega}}\left[a_k\phi_k(x)+a_k^\dagger\phi_k^*(x)\right]
\end{split}
\end{equation}
where we have used the WKB condition $\lambda\approx k_z$. $\phi_k$ is defined in Eq.(\ref{solutionform}) with the complex orthogonal relation $\int d^3\vec{x}\phi_k(x)\phi^*_p(x)=(2\pi)^3\delta^3(k-p)$, $\int d^3\vec{p}/(2\pi)^3 \phi_k(x)\phi_k^*(y)=\delta^3(x-y)$. Then, the conjugate momentum can be expressed by
\begin{equation}
    \hat{\pi}(x)\approx-i\int\frac{d^3\vec{k}}{(2\pi)^3}\sqrt{\frac{\omega}{2}}\left[a_k\phi_k(x)-a_k^\dagger\phi_k^*(x)\right]
\end{equation}
multiply field and conjugate momentum by $\phi_p^*(x)$ and integrate out space-coordinate we have
\begin{equation}
\begin{split}
    \int d^3\vec{x}\hat{\phi}(x)\phi_p^*(x)=\int d^3\vec{x}\frac{d^3\vec{k}}{(2\pi)^3}\frac{1}{\sqrt{2\omega}}[a_p\phi_k(x)\phi_p^*(x)+a_k^\dagger\phi_k^*(x)\phi_p^*(x)]
    \\
    \int d^3\vec{x}\hat{\pi}(x)\phi_p^*(x)=-i\int d^3\vec{x}\frac{d^3\vec{k}}{(2\pi)^3}\sqrt{\frac{\omega}{2}}[a_p\phi_k(x)\phi_p^*(x)-a_k^\dagger\phi_k^*(x)\phi_p^*(x)]
\end{split}
\end{equation}
Using the orthogonal relationship of $\phi_k(x)$, one can prove that the creation and annihilation operators can be expressed by
\begin{equation}
\begin{split}
    a_p=\int d^3\vec{x}\frac{1}{\sqrt{2\omega}}[\omega\hat{\phi}(x)+i\hat{\pi}(x)]\phi_p^*(x)
    \\
    a_p^\dagger=\int d^3\vec{x}\frac{1}{\sqrt{2\omega}}[\omega\hat{\phi}(x)-i\hat{\pi}(x)]\phi_p(x)
\end{split}
\end{equation}
then one can quantize the field following the canonical quantization
\begin{equation}
    [\hat{\phi}(t,\vec{x}),\hat{\pi}(t,\vec{y})]=i\delta^3(x-y)\quad\quad [\hat{\phi}(t,\vec{x}),\hat{\phi}(t,\vec{y})]=[\hat{\pi}(t,\vec{x}),\hat{\pi}(t,\vec{y})]=0
\end{equation}
then one can get the commutation relation of the creation and annihilation operators is given by
\begin{equation}
\begin{split}
    [a_k,a_p^\dagger]&=\int d^3\vec{x}\int d^3\vec{y}\frac{1}{2\omega}[\omega\hat{\phi}(x)+i\hat{\pi}(x),\omega\hat{\phi}(y)+i\hat{\pi}(y)]\phi^*_k(x)\phi_p(y)
    \\
    &=\int d^3\vec{x} \phi_k^*(x)\phi_p(x)=(2\pi)^3\delta(k-p)
\end{split}
\end{equation}
the another two condition $[a_k,a_p]=[a_k^\dagger,a_p^\dagger]=0$ can be proved by same method. One can also verify the commutation relation of field and conjugate momentum
\begin{equation}
\begin{split}
    [\hat{\phi}(t,\vec{x}),\hat{\pi}(t,\vec{y})]&=\frac{-i}{2}\int\frac{d^3\vec{k}}{(2\pi)^3}\int\frac{d^3\vec{p}}{(2\pi)^3}[a_k\phi_k(x)+a_k^\dagger\phi_k^*(x),a_p\phi_p(x)-a_p^\dagger\phi_p^*(x)]
    \\
    &=\frac{i}{2}\int\frac{d^3\vec{k}}{(2\pi)^3}[\phi_k^*(x)\phi_k(y)+\phi_k(x)\phi_k^*(y)]=i\delta^3(x-y)
\end{split}
\end{equation}
Here, we have successfully quantized this field. 

Then, we need to discuss the external line of each field; the scalar has non-external line, and the vector external line is the polarization vector. So, we find the WKB solution for scalar and vector fields. Now, let us discuss the solution for the Dirac field. For an ordinary Dirac field, the solution is given by $\psi(x)=u(k)e^{-ik\cdot x}$. The structure of this solution is given by the Dirac spinor produced by the solution of the scalar field. So, we assume the Dirac field with $z$ dependent mass has the same structure.
\begin{equation}
    \psi(x)=\tilde{u}(k)\phi(x)=u'(k)\chi_k(z)e^{-i(Et-\vec{k}_\perp\cdot\vec{x}_\perp)},
\end{equation}
where $\phi(x)$ and $\chi(z)$ is solution for scalar field and $u'(p)$ is undetermined spinor function. Taking this solution into EoM, one can find the equation of $u'(p)$ is given by
\begin{equation}
        [-\gamma_0 E\chi_k+\gamma_\perp\cdot k_\perp\chi_k-i\gamma_z \frac{d\chi_k(z)}{dz}+m(z)]u'(p)=0.
\end{equation}
Using the solution of $\chi(z)$, one will find $d\chi(z)/dz=[ik_z(z)-k_z'(z)/(2k_z(z))]\chi(z)$. With the WKB condition condition $k_z(z)\gg k_z'(z)/k_z(z)$, one can find the equation of $u'(p)$ is given by $[\gamma^\mu p_\mu-m(z)]u'(p)=0$. This equation is the same equation for the ordinary Dirac spinor function. So, we find $u'(p)=u(p)$. The next step is to write the field operator using those solutions. This is actually already done by \cite{Azatov:2023xem,Kubota:2024wgx} by considering the reflection/transmission effect and left(from symmetry phase to Higgs phase)/right(from Higgs phase to symmetry phase) moving particles. However, for the problems we want to consider (ultra fast-moving wall), the reflection and right moving particles effect can be ignored 
In this case, the field operator can be expanded as
\begin{equation}
\begin{split}
    &\hat{\phi}\approx\int\frac{d^3\vec{k}}{(2\pi)^3}\frac{1}{\sqrt{2E_k}}[\hat{a}_k\chi_k(z)e^{-i(Et-k_\perp\cdot x_\perp)}+\hat{a}^\dagger_k\chi_k^{*}(z)e^{i(Et-k_\perp\cdot x_\perp)}]
    \\
    &\hat{A}_\mu\approx\int\frac{d^3\vec{k}}{(2\pi)^3}\frac{1}{\sqrt{2E_k}}\sum_{s}[\hat{a}^s_k\epsilon^s_\mu(p)\chi_k(z)e^{-i(Et-k_\perp\cdot x_\perp)}+\hat{a}^{s,\dagger}_k\epsilon^{s,*}_\mu(k)\chi_k^{*}(z)e^{i(Et-k_\perp\cdot x_\perp)}]
    \\
    &\hat{\psi}\approx\int\frac{d^3\vec{k}}{(2\pi)^3}\frac{1}{\sqrt{2E_k}}\sum_{s}[\hat{a}^s_k u^s(k)\chi_k(z)e^{-i(Et-k_\perp\cdot x_\perp)}+\hat{a}^{s,\dagger}_k v^s(k)\chi_k^{*}(z)e^{i(Et-k_\perp\cdot x_\perp)}],
\end{split}
\end{equation}
with $[a_k,a_p^\dagger]=(2\pi)^3\delta^3(k-p)$ and $[a_k,a_p]=[a_k^\dagger,a_p^\dagger]=0$. 

Combining all the discussions, we can formulate the Feynman rules for the external line as
\begin{equation}
\begin{split}
   &\wick{\c \phi_x\vert \c p}\rangle=\chi_p(z)e^{-i(Et-\vec{p}_\perp\cdot\vec{x}_\perp)}
    \ \ \ \ \ \ \ \ \ \ \ \ \ 
   \langle\wick{\c p \vert \c\phi_x}=\chi_p^*(z)e^{i(Et-\vec{p}_\perp\cdot\vec{x}_\perp)}
   \\
   &\wick{\c \phi^\dagger_x\vert \c p}\rangle=\chi_p^*(z)e^{i(Et-\vec{p}_\perp\cdot\vec{x}_\perp)}
    \ \ \ \ \ \ \ \ \ \ \ \ \ \ \ 
   \langle\wick{\c p \vert \c\phi^\dagger_x}=\chi_p(z)e^{-i(Et-\vec{p}_\perp\cdot\vec{x}_\perp)}
   \\
   &\wick{\c A_x^\mu \vert \c p}\rangle=\epsilon_\mu(p)\chi_p(z)e^{-i(Et-\vec{p}_\perp\cdot\vec{x}_\perp)}
    \ \ \ \ \ \ 
   \langle\wick{\c p \vert \c A^\mu_x}=\epsilon^*_\mu(p)\chi_p^*(z)e^{i(Et-\vec{p}_\perp\cdot\vec{x}_\perp)}
   \\
   &\wick{    \c\psi_x \vert \c p}  \rangle=u(p)\chi_p(z)e^{-i(Et-\vec{p}_\perp\cdot\vec{x}_\perp)}
    \ \ \ \ \ \ \ 
     \langle \wick{  \c p \vert \c\psi_x }=v(p)\chi_p^*(z)e^{i(Et-\vec{p}_\perp\cdot\vec{x}_\perp)}
   \\     
   &\wick{    \c{\bar{\psi}}_x  \vert\c  p }  \rangle=\bar{v}(p)\chi(z)e^{-i(Et-\vec{p}_\perp\cdot\vec{x}_\perp)}
    \ \ \ \ \ \ \ \ 
   \langle\wick{  \c p \vert \c{\bar{ \psi}}_x }=\bar{u}(p)\chi_p^*(z)e^{i(Et-\vec{p}_\perp\cdot\vec{x}_\perp)}.
\end{split}
\end{equation}

\section{The friction pressure from the Splitting effect}\label{splitting}
The friction force from the splitting effect is given by
\begin{equation}
\begin{split}
    \frac{F^\mathrm{fric}_{a\rightarrow bc}}{A}&=\int\frac{d^3\vec{p}_a}{(2\pi)^3}\frac{p_a^z}{E_a}f_a\int dP_{a\rightarrow bc}(p_a^z-p_b^z-p_c^z),
\end{split}
\end{equation}
The key to computing the friction force is the probability, which is given by
\begin{equation}
    \int dP_{a\rightarrow bc}=\int\frac{d^3\vec{p}_b}{(2\pi)^3}\frac{1}{2E_b}\int\frac{d^3\vec{p}_c}{(2\pi)^3}\frac{1}{2E_c}\langle \phi|i\mathcal{T}|p_b,p_c\rangle\langle p_b,p_c|i\mathcal{T}|\phi\rangle,
\end{equation}
where the incoming state $|\phi\rangle$ is defined by
\begin{equation}
    |\phi\rangle=\int\frac{d^3\vec{p}}{(2\pi)^3}\frac{\phi(p)}{2E_p}|p\rangle,\ \ {\rm with }\ \ \int\frac{d^3\vec{p}}{(2\pi)^3}\frac{|\phi(p)|^2}{2E_p}=1,
\end{equation}
For single color wave we have $|\phi(p)|^2=2E_p(2\pi)^3\delta^{3}(\vec{p}-\vec{p}_a)$. So, the probability can be written by
\begin{equation}
    \int dP_{a\rightarrow bc}=\int\frac{d^3\vec{p}_b}{(2\pi)^3}\frac{1}{2E_b}\int\frac{d^3\vec{p}_c}{(2\pi)^3}\frac{1}{2E_c}\int\frac{d^3\vec{p}}{(2\pi)^3}\frac{\phi(p)}{2E_p}\int\frac{d^3\vec{p}'}{(2\pi)^3}\frac{\phi^*(p')}{2E_{p'}}\langle p'|i\mathcal{T}|p_b,p_c\rangle\langle p_b,p_c|i\mathcal{T}|p\rangle,
\end{equation}
the $\mathcal{M}$-matrix element can only be determined by interaction. For QED-type interaction, the matrix element can be written by
\begin{equation}\label{appd:amplitude}
    \langle p|iT|p_b,p_c\rangle=-ig\int d^4x \langle p|\bar{\psi}_a(x)\gamma^\mu\psi_b(x)A_\mu(x)|p_b,p_c\rangle,
\end{equation}
using the contraction rule from Section.\ref{BQFT}, we can simplify this matrix element as
\begin{equation}\label{E4}
    \langle p|iT|k_b,k_c\rangle=\int d^4x \chi^*_p(z)V(z)\chi_b(z)\chi_c(z)e^{i(E_a-E_b-E_c)t}e^{-i(\vec{p}_a-\vec{p}_b-\vec{p}_c)_\perp\cdot\vec{x}_\perp},
\end{equation}
where $V(z)=-ig\bar{u}(p)\gamma^\mu u(k_b)\epsilon_\mu(k_c)$. Integrating out all the coordinates except $z$, we will find that
\begin{equation}\label{app:amp}
    \langle p|iT|k_b,k_c\rangle=M_{a\rightarrow bc}(2\pi)^3\delta(E_a-E_b-E_c)\delta^2(\vec{p}^\perp-\vec{p}^\perp_b-\vec{p}^\perp_c),
\end{equation}
where $M_{a\rightarrow bc}=\int dz \chi^*_p(z)V(z)\chi_b(z)\chi_c(z)$. Then, substituting this element into the expression of probability, one will find that
\begin{equation}
\begin{split}
    \int dP_{a\rightarrow bc}&=\int\frac{d^3\vec{p}_b}{(2\pi)^3}\frac{1}{2E_b}\int\frac{d^3\vec{p}_c}{(2\pi)^3}\frac{1}{2E_c}\int\frac{d^3\vec{p}}{(2\pi)^3}\frac{\phi(p)}{2E_p}\int\frac{d^3\vec{p}'}{(2\pi)^3}\frac{\phi^*(p')}{2E_{p'}}|M|^2_{a\rightarrow bc}
    \\
    &(2\pi)^3\delta(E_p-E_b-E_c)\delta^2(\vec{p}^\perp-\vec{p}^\perp_b-\vec{p}^\perp_c)(2\pi)^3\delta(E_{p'}-E_b-E_c)\delta^2(\vec{p}'^\perp-\vec{p}^\perp_b-\vec{p}^\perp_c)
    \\
    &=\int\frac{d^3\vec{p}_b}{(2\pi)^3}\frac{1}{2E_b}\int\frac{d^3\vec{p}_c}{(2\pi)^3}\frac{1}{2E_c}\int\frac{d^3\vec{p}}{(2\pi)^3}\frac{|\phi(p)|^2}{2E_p}\frac{1}{2p^z}|M|^2_{a\rightarrow bc}
    \\
    &\ \ \ \ \ \ \ \ \ \ \ \ \ \ \ \ \ \ \ \ \ \ \ \ \ \ \ \ \ \ \ \ \ \ \ \ \ \ \ \ (2\pi)^3\delta(E_p-E_b-E_c)\delta^2(\vec{p}^\perp-\vec{p}^\perp_b-\vec{p}^\perp_c)
    \\
    &=\int\frac{d^3\vec{k}_b}{(2\pi)^3}\frac{1}{2E_b}\int\frac{d^3\vec{k}_c}{(2\pi)^3}\frac{1}{2E_c}\frac{1}{2p_a^z}|M|^2_{a\rightarrow bc}(2\pi)^3\delta(E_a-E_b-E_c)\delta^2(\vec{p}_a^\perp-\vec{k}^\perp_b-\vec{k}^\perp_c),
\end{split}
\end{equation}
Using this probability, one can rewrite this friction force as as
\begin{equation}
\begin{split}
    \frac{F^\mathrm{fric}_{a\rightarrow bc}}{A}=\int\frac{d^3\vec{p}_a}{(2\pi)^3}\frac{f_a}{2E_a}\int&\frac{d^3\vec{p}_b}{(2\pi)^3}\frac{d^3\vec{p}_c}{(2\pi)^3}\frac{|M|^2_{a\rightarrow bc}}{2E_b 2E_c}(p_a^z-p_b^z-p_c^z)
    \\
    &\times(2\pi)^3\delta(E_a-E_b-E_c)\delta^2(\vec{p}_a^\perp-\vec{p}^\perp_b-\vec{p}^\perp_c).
\end{split}
\end{equation}

Now, using the above definition and a similar derivation process, we can write the Boltzmann equation, involving $a\rightarrow b c$ splitting process, in a more accurate form. To do that, we can first write the Boltzmann equation involving splitting as
\begin{equation}\label{BES}
    \frac{p_z}{E_p}\frac{\partial}{\partial z}f_a(p,z)+\frac{\partial}{\partial p_z}[-\frac{d m_a^2}{d z}\frac{f_a(p,z)}{E_p}]=\mathcal{C}[f_a]
\end{equation}
and the collision terms with $a\rightarrow b c$ splitting process can be written as
\begin{equation}\label{con:sp}
\begin{split}
    \mathcal{C}[f_a]=-\int\frac{d^3\vec{p}_b}{(2\pi)^3}\int\frac{d^3\vec{p}_c}{(2\pi)^3}\frac{f_a(x,p)|\langle b,c|i\mathcal{T}|a\rangle|^2}{TV2E_a}\frac{[1\pm f_b(x,p_b)]}{2E_b}\frac{[1\pm f_c(x,p_c)]}{2E_c}&
    \\
    +{\rm Inverse\ Process}&,
\end{split}
\end{equation}
where we have used the fact that the bubble is propagating along the $z$-direction. As a consequence, $\vec{F}$ vanished except for $F_z$ and the $f_a$ only depended on the $z$ coordinate. Then, let us rewrite the amplitude square through the Eq.\eqref{app:amp} and we can find
\begin{equation}\label{BOX}
\begin{split}
    \frac{|\langle b,c|i\mathcal{T}|a\rangle|^2}{TV}&=|M_{a\rightarrow bc}|^2\delta^{(3)}(p_a^\perp-p_b^\perp-p^\perp_c)\frac{(2\pi)^3\delta^{(3)}(0)}{TL^3}
    \\
    &=\frac{|M_{a\rightarrow bc}|^2}{L}(2\pi)^3\delta^{(3)}(p_a^\perp-p_b^\perp-p^\perp_c)
\end{split}
\end{equation}
where we have used the Box Normalization condition $2\pi\delta(0)=L$ or $T$ and $V=L^3$. And finally, combine Eq.\eqref{BES}, Eq.\eqref{con:sp} amd Eq.\eqref{BOX}, we can written down the Boltzmann equation in BQFT method as
\begin{equation}
\begin{split}
    \bigg[k_z&\frac{\partial}{\partial z}-\frac{dm^2(z)}{dz}\frac{\partial}{\partial k_z} \bigg]\frac{f_a(k,z)}{2E_k}=
    \\
    &-\int\frac{d^3\vec{p}_b}{(2\pi)^3}\int\frac{d^3\vec{p}_c}{(2\pi)^3}\frac{f_a(k,z)|M_{a\rightarrow bc}|^2}{2E_aL}\frac{1\pm f_b(p_b,z)}{2E_b}\frac{1\pm f_c(p_c,z)}{2E_c}
    \\
    &\quad\quad\quad\times(2\pi)^3\delta(E_k-E_p-E_{p'})\delta^2(\vec{k}_\perp-\vec{p}_\perp-\vec{p}'_\perp)+{\rm Inverse\ Process}.
\end{split}
\end{equation}

\section{The Kadanoff-Baym Equation}\label{driveofKBE}
In this section, we will give a basic introduction to the Kadanoff-Baym equation. For the scalar field theory, the Green's function in Closed Time Path Formalism can be written as
\begin{align}
\wt{G}(x,y)=&
\begin{pmatrix}
    G^t(x,y) & -G^<(x,y)\\
    G^>(x,y) & -G^{\bar{t}}(x,y)
\end{pmatrix}
\end{align}
where the component is defined as
\begin{equation}\label{CTPG}
\begin{split}
    &G^>(x,y)=\langle\phi(x)\phi^\dagger(y)\rangle\\
    &G^<(x,y)=\langle\phi^\dagger(y)\phi(x)\rangle\\
    &G^t(x,y)=\Theta(x_0-y_0)G^>(x,y)+\Theta(y_0-x_0)G^<(x,y)\\
    &G^{\bar{t}}(x,y)=\Theta(x_0-y_0)G^<(x,y)+\Theta(y_0-x_0)G^>(x,y)
\end{split}
\end{equation}
The EOM of Green's function are given by the 2-point Dyson-Schwinger equation. Here we have
\begin{equation}\label{DSEG}
\begin{split}
    [\partial_x^2+m^2(x)]\wt G(x,y)=-i\delta^4(x-y)\wt I-i\int d^4z\wt \Pi(x,z)\wt G(z,y)
    \\
    \wt G(x,y)[\overleftarrow{\partial}_y^2+m^2(y)]=-i\delta^4(x-y)\wt I-i\int d^4z\wt G(x,z)\wt \Pi(z,y)
\end{split}
\end{equation}
where $\wt\Pi$ is the self-energy matrix, which can be defined by Eq.(\ref{CTPG}), similarly. Now, we can use those equations to write down the kinetic equation for Green's function. To do that, we need to use the Wigner transformation Eq.(\ref{wigner}) in the above equation. We can write Eq.(\ref{DSEG}) by the average coordinate and relative coordinate by $x=X+\frac{r}{2}$, $y=X-\frac{r}{2}$. We can obtain
\begin{equation}
\begin{split}
    &\bigg[\frac{1}{4}\partial_X^2+\partial_r^2+\partial_X\cdot\partial_r-m^2(x+\frac{r}{2})\bigg]\wt G(x,y)=-i\delta^4(r)-i\int d^4z \wt\Pi(x,z)\wt G(z,y)
    \\
    &\wt G(x,y)\bigg[\frac{1}{4}\overleftarrow{\partial}_X^2+\overleftarrow{\partial}_r^2+\overleftarrow{\partial}_X\cdot\overleftarrow{\partial}_r-m^2(x+\frac{r}{2})\bigg]=-i\delta^4(r)-i\int d^4z \wt G(x,z)\wt\Pi(z,y),
\end{split}
\end{equation}
Then, take the Wigner transformation of the sum and difference, respectively, the above equation would become
 \begin{equation}
 \begin{split}
     &\bigg(\frac{1}{2}\partial_X^2-2k^2\bigg)\wt G(k,X)+e^{-i\Diamond}\{m^2(X),\wt G(k,X)\}=-2i\wt I-ie^{-i\Diamond}{\wt\Pi(k,X),\wt G(k,X)}
     \\
     &-2ik\cdot\partial_X\wt G(k,X)+e^{-i\Diamond}[m^2(X),\wt G(k,X)]=-ie^{-i\Diamond}[\wt\Pi(k,X),\wt G(k,X)]
 \end{split}
 \end{equation}
 where we have used the property that the Wigner transformation of the spacetime convolution can be written as
\begin{equation}
    \int d^4r e^{ik\cdot r}\int d^4z A(x,z)B(z,y)=A(k+\frac{i}{2}\partial_X,X-\frac{i}{2}\partial_k)B(k,X)=e^{-i\Diamond}(A(k,X)B(k,X))
\end{equation}
The Kadanoff-Baym equation is just this matrix equation's $<$ and $>$ components.

\section{The constrain equation}\label{APPE}
In this section, we will give the legitimacy of Approximation-2. Let us set the coupling coordinate independent $Y(x)=\lambda$, for simplicity. The constrain equation for the Kadanoff-Baym equation can be written as
\begin{equation}
\begin{split}
    (\frac{1}{2}\partial_X^2-2k^2)&G^<(k,z)+G^<(k-\frac{i}{2}d_z,z+\frac{i}{2}\partial_{k_z})m^2(z)+G^<(k+\frac{i}{2}d_z,z-\frac{i}{2}\partial_{k_z})m^2(z)
    \\
    &=\frac{iY^2}{4}\int\frac{d^4p}{(2\pi)^4}\frac{d^4p'}{(2\pi)^4}(2\pi)^4\delta^4(k-\frac{i}{2}d_z-p-p')
    \\
    &\quad\quad\quad G_\Phi^>(p,z-\frac{i}{2}\partial_{k_z}) G_\Phi^>(p',z-\frac{i}{2}\partial_{k_z})G_\phi^<(k,z)-(d_z\rightarrow-d_z)
    \\
    &+(<\leftrightarrow >)
\end{split}
\end{equation}
where we have ignored the contribution from $\Pi_h$ and $G_h$. Then expand the Green's function by the $\epsilon_{wall}$ and only keep the leading order. We have
\begin{equation}\label{conKB}
\begin{split}
    -2(k^2-&m^2(z))G^<(k,z)
    \\
    &=\frac{iY^2}{4}\int\frac{d^4p}{(2\pi)^4}\frac{d^4p'}{(2\pi)^4}(2\pi)^4\delta^4(k-\frac{i}{2}d_z-p-p')
    \\
    &\quad\quad\quad G_\Phi^>(p,z) G_\Phi^>(p',z)G_\phi^<(k,z)-(d_z\rightarrow-d_z)
    \\
    &+(<\leftrightarrow >)
\end{split}
\end{equation}
To deal with this equation, we can rewrite the Green's function $G^<$ by the Wigner transformation Eq.(\ref{NWigner}) twice
\begin{equation}
    G^<(k,z)=\int dz' \int\frac{dl_z}{2\pi}G^<(k,z')e^{il_z(z-z')}
\end{equation}
then substituting this equation into Eq.(\ref{conKB}), then we got
\begin{equation}
\begin{split}
    -2(k^2-&m^2(z))G^<(k,z)
    \\
    &=\frac{i Y^2}{4}\int\frac{d^4p}{(2\pi)^3}\frac{d^4p'}{(2\pi)^4}(2\pi)^3\delta(k_0-p_0-p'_0)\delta^2(\vec{k}_\perp-\vec{p}_\perp-\vec{p}'_\perp)
    \\
    &\quad\quad\quad G_\Phi^>(p,z) G_\Phi^>(p',z)\int dz' G^<(k,z') e^{-2i\Delta p_z (z-z')}-(\Delta p_z\rightarrow-\Delta p_z)
    \\
    &+(<\leftrightarrow >).
\end{split}
\end{equation}
If the bubble wall velocity is very large, then the energy of the particle will dominate compared with the mass change, which is given by the background field $\gamma T>>\Delta m$. In this case, $G^<(k,z')$ will be weak dependence on the coordinate $z'$ and we have 
\begin{equation}
\begin{split}
    \int dz' G^<(k,z') e^{-2i\Delta p_z (z-z')}-(\Delta p_z\rightarrow-\Delta p_z)&\approx G^<(k) \Bigg(\int dz' e^{il_z(z-z')}-\int dz' e^{il_z(z'-z)}\Bigg)
    \\
    &\approx 0
\end{split}
\end{equation}
then the constrain equation will become
\begin{equation}
    [k^2-m^2(z)]G^\gtrless(k,z)\approx 0
\end{equation}
which is just the ordinary LO constraint equation for the Kadanoff-Baym equation. The general solution for this equation is just the Eq.(\ref{SCKBE}). That is to say, Approximation-2 is reasonable.

\section{Computation of the friction pressure for $\phi^2\leftrightarrow\Phi^2$ process}\label{App:Com}

In this appendix, we will show how we simplified the friction pressure in $\phi^2\leftrightarrow\Phi^2$ the scattering process and obtain the Eq.\eqref{fric22c3}.  Firstly, we can follow the logic below Eq.\eqref{MMphi} and do the approximation. After the approximation, the amplitude for $\phi^2\rightarrow\Phi^2$ and the inverse process in Eq.(\ref{MMphi}) became:
\begin{equation}\label{amsimp}
\begin{split}
    &M_{\phi^2\rightarrow\Phi^2}\approx 2i\lambda\frac{E_kE_k'}{E_p+E_p'}\frac{\Delta m_\Phi^2}{(\vec{k}_\perp^2+{m^s_\Phi}^2)(\vec{k}_\perp^2+{m^h_\Phi}^2)},
    \\
    &M_{\Phi^2\rightarrow\phi^2}\approx
       \left\{ 
       \begin{array}{l} 
        \frac{2i\lambda E_pE_p'}{E_p+E_p'}\frac{\Delta m_\Phi^2}{{m_\Phi^h}^2{m_\Phi^s}^2}\quad\quad\quad\quad\quad\quad\quad\quad\quad\vec{k}_\perp^2+{m_\phi^h}^2<{m_\Phi^h}^2, \\ 
        2i\lambda\frac{E_kE_k'}{E_p+E_p'}\frac{\Delta m_\phi^2}{(\vec{k}_\perp^2+{m_\phi^s}^2)(\vec{k}_\perp^2+{m_\phi^h}^2)}\quad\quad\quad\quad\vec{k}_\perp^2+{m_\phi^h}^2>{m_\Phi^h}^2. 
       \end{array} \right.
       \\
    &M_{\phi\Phi\rightarrow\phi\Phi}\approx2i\lambda\bigg[\frac{E_k E_k'}{\vec{k}_\perp^2(E_p+E_p')+{m_\phi^s}^2E_k'+{m_\Phi^s}^2E_k}-\frac{E_k E_k'}{\vec{k}_\perp^2(E_p+E_p')+{m_\phi^h}^2E_k'+{m_\Phi^h}^2E_k}\bigg]
    \\
    &\quad\quad\quad\ \ \approx2i\lambda\frac{E_k E_k'}{\vec{k}_\perp^2(E_p+E_p')+{m_\phi^s}^2E_k'+{m_\Phi^s}^2E_k}.
\end{split}
\end{equation}
where, for the scattering process, we have used the fact that $m^h_i\gg m^s_i$, and ignored the subdominant contribution. The corresponding momentum transfer can also be simplified by using those approximation
\begin{equation}\label{montrans}
    p_z+p_z'-k_z-k_z'\approx
       \left\{ 
       \begin{array}{l} 
        \frac{(E_k+E_k')(\vec{k}_\perp^2+{m^h_{\Phi}}^2)}{4E_kE_k'}\quad\quad\quad\quad\quad\quad\ \quad\quad\quad\quad\quad\quad\phi^2\rightarrow\Phi^2, \\ 
        \frac{(E_k+E_k')(\vec{k}_\perp^2+{m^h_{\phi}}^2)}{4E_kE_k'}-\frac{(E_p+E_p'){m^s_{\Phi}}^2}{4E_pE_p'}\quad\quad\quad\quad\quad\quad\ \Phi^2\rightarrow\phi^2,\\
        \frac{\vec{k}_\perp^2(E_k+E_k')+{m_\phi^h}^2E_k'+{m_\Phi^h}^2E_k}{2E_k E_k'}-\frac{{m_\phi^s}^2E_p'+{m_\Phi^s}^2E_p}{2E_p E_p'}
        \\
        \quad\quad\quad\quad\quad\ \approx\frac{\vec{k}_\perp^2(E_k+E_k')+{m_\phi^h}^2E_k'+{m_\Phi^h}^2E_k}{2E_k E_k'}\quad\quad\phi\Phi\rightarrow\phi\Phi.
       \end{array} \right.
\end{equation}
where we used the energy conservation relation $E_p+E_p'-E_k-E_k'=0$ and, for the scattering process, we still used the fact that $m^h_i\gg m^s_i$. Substituting this expression into Eq.(\ref{pre22}), and use the high-wall velocity approximation which cause $E_p'\approx p_z'\gg k_\perp,m_\Phi$ for all the particles. 

Following the logic in Section.\ref{22preV}, we can substituting Eq.\eqref{amsimp} and Eq.\eqref{montrans} into Eq.\eqref{pre22} and obtain
\begin{equation}\label{fric22cc}
\begin{split}
    \frac{F_{\phi^2\leftrightarrow\Phi^2}^{fric}}{A}&\approx
    \int\frac{d^3\vec{p}}{(2\pi)^3}\int\frac{d^3\vec{p}'}{(2\pi)^3}\frac{f_{\phi}(p)f_{\phi}(p')}{E_pE_p'(E_p+E_p')}\int\frac{d^3\vec{k}}{(2\pi)^3}\frac{\lambda^2\Delta m_\Phi^4(\vec{k}_\perp^2+ m_\Phi^2)}{16(\vec{k}_\perp^2+{m_\Phi^s}^2)^2(\vec{k}_\perp^2+{m_\Phi^h}^2)^2}
    \\
    &+\int\frac{d^3\vec{p}}{(2\pi)^3}\int\frac{d^3\vec{p}'}{(2\pi)^3}\frac{f_{\Phi}(p)f_{\Phi}(p')}{E_p E_p'(E_p+E_p')}\int\frac{d^3\veck}{(2\pi)^3}\frac{\lambda^2\Delta m_\phi^4(\veck_\perp^2+{m_\phi^h}^2)}{16(\vec{k}_\perp^2+{m_\phi^s}^2)^2(\vec{k}_\perp^2+{m_\phi^h}^2)^2}|_{\veck_\perp^2+m_\phi^2>m_\Phi^2}
    \\
    &-\int\frac{d^3\vec{p}}{(2\pi)^3}\int\frac{d^3\vec{p}'}{(2\pi)^3}\frac{f_{\Phi}(p)f_{\Phi}(p')}{E_p^2E_p'^2(E_p+E_p')}\int\frac{d^3\veck}{(2\pi)^3}\frac{E_k E_k'\Delta m_\phi^4 {m_\Phi^s}^2}{16(\vec{k}_\perp^2+{m_\phi^s}^2)^2(\vec{k}_\perp^2+{m_\phi^h}^2)^2}|_{\veck_\perp^2+m_\phi^2>m_\Phi^2}
    \\
    &+\int\frac{d^3\vec{p}}{(2\pi)^3}f_{\Phi}(p)\int\frac{d^3\vec{p}'}{(2\pi)^3}f_{\Phi}(p')\frac{E_pE_p'}{E_p+E_p'}\int\frac{d^3\vec{k}}{(2\pi)^3}\frac{\lambda^2\Delta m_\Phi^4}{E_k^2E_k'^2}\frac{\vec{k}_\perp^2+ m_\phi^2}{16{m_\Phi^h}^4{m_\Phi^s}^4}|_{\vec{k}^2_\perp+m_\phi^2<m_\Phi^2}
    \\
    &-\int\frac{d^3\vec{p}}{(2\pi)^3}f_{\Phi}(p)\int\frac{d^3\vec{p}'}{(2\pi)^3}f_{\Phi}(p')\frac{1}{E_p+E_p'}\int\frac{d^3\vec{k}}{(2\pi)^3}\frac{\lambda^2\Delta m_\Phi^4}{E_kE_k'}\frac{1}{16{m_\Phi^h}^4{m_\Phi^s}^2}|_{\vec{k}^2_\perp+m_\phi^2<m_\Phi^2}\nonumber
\end{split}
\end{equation}
\begin{equation}\label{eq:firctotal}
\quad\quad\quad+\int\frac{d^3\vec{p}}{(2\pi)^3}\frac{f_{\Phi}(p)}{2E_p}\int\frac{d^3\vec{p}'}{(2\pi)^3}\frac{f_{\Phi}(p')}{2E_p'}\int\frac{d^3\vec{k}}{(2\pi)^3}\frac{\lambda^2}{2}\frac{\vec{k}_\perp^2(E_p+E_p')+{m^h_\Phi}^2E_k'+{m^h_\phi}^2E_k}{[\vec{k}_\perp^2(E_p+E_p')+{m^s_\Phi}^2E_k'+{m^s_\phi}^2E_k]^2}
\end{equation}
The first line represents the friction force that comes from the process $\phi^2\rightarrow\Phi^2$. The second and third lines represent the contribution from the inverse process $\Phi^2\rightarrow\phi^2$ with $\veck_\perp^2+m_\phi^2>m_\Phi^2$. The fifth and sixth lines represent the contribution from the inverse process $\Phi^2\rightarrow\phi^2$ but with $\veck_\perp^2+m_\phi^2<m_\Phi^2$. In principle, we should evaluate all terms in this friction pressure, but it is not necessary. 
One can directly drop the contribution from the second and third lines by observing that the contribution from the second and third has the same structure as the contribution from the firstly but is smaller since $m_\Phi^2>m_\phi^2$. So, the total friction pressure from $\phi^2\rightarrow\Phi^2$ in the phase transition can be written as
\begin{equation}\label{fric22cc}
\begin{split}
    \frac{F_{\phi^2\rightarrow\Phi^2}^{fric}}{A}\approx A_1(\gamma,T)\times A_2+I_1(\gamma,T)\times I_2&-T_1(\gamma,T)\times T_2+P_s(\gamma,T).
\end{split}
\end{equation}
where the $R_1,I_1,T_1$ are the $\gamma_w$ dependent function and $R_2,I_2,T_2$ are the constant coefficient:
\begin{align}
    &A_1(\gamma,T)=\int \frac{d^3\vec{p}}{(2\pi)^3}\int \frac{d^3\vec{p}'}{(2\pi)^3}\int \frac{dk_z}{2\pi}\frac{f(p)f(p')}{E_pE_p'(E_p+E_p')}\label{G5}
    \\
    &A_2=\int\frac{d^2\vec{k}_\perp}{(2\pi)^2}\frac{\lambda^2\Delta m_\Phi^4(\vec{k}_\perp^2+\Delta m_\Phi^2)}{16(\vec{k}_\perp^2+{m_\Phi^s}^2)^2(\vec{k}_\perp^2+{m_\Phi^h}^2)^2}
    \\
    &I_1(\gamma,T)=\int \frac{d^3\vec{p}}{(2\pi)^3}\int \frac{d^3\vec{p}'}{(2\pi)^3}\int \frac{dk_z}{2\pi} \frac{p_z p_z'f(p)f(p')}{(p_z+p_z')(k_z^2+{m_\phi^h}^2)(p_z+p_z'-\sqrt{k_z^2+{m^h_\phi}^2})^2}
    \\
    &I_2=\int\frac{d^2\vec{k}_\perp}{(2\pi)^2}\frac{\lambda^2\Delta m_\Phi^4(\vec{k}_\perp^2+ m_\phi^2)}{16{m_\Phi^h}^4{m_\Phi^s}^4}\Theta(m_\Phi^2-\vec{k}^2_\perp-m_\phi^2)
\end{align}
\begin{align}
    &T_1(\gamma,T)=\int \frac{d^3\vec{p}}{(2\pi)^3}\int \frac{d^3\vec{p}'}{(2\pi)^3}\int \frac{dk_z}{2\pi} \frac{f(p)f(p')}{(p_z+p_z')\sqrt{k_z^2+{m_\phi^h}^2}(p_z+p_z'-\sqrt{k_z^2+{m^h_\phi}^2})}
    \\
    &T_2=\int\frac{d^2\vec{k}_\perp}{(2\pi)^2}\frac{\lambda^2\Delta m_\Phi^4}{16{m_\Phi^h}^4{m_\Phi^s}^2}\Theta(m_\Phi^2-\vec{k}^2_\perp-m_\phi^2)
\end{align}
where we have ignored the $m^s$ in the energy of incoming particles for simplicity. The integral area is to make $E_k,E_k'>0$, which is $p_z,p_z'>m_\Phi$, $0<k_z<\sqrt{(p_z+p_z')^2-m_\Phi^2}$ and $\vec{k}_\perp^2>0$. The $P_s(\gamma,T)$ is the last friction force in Eq.\eqref{eq:firctotal} contributing by the scattering process.
\begin{figure}[htbp] 
    \centering
    \includegraphics[width=0.45\textwidth]{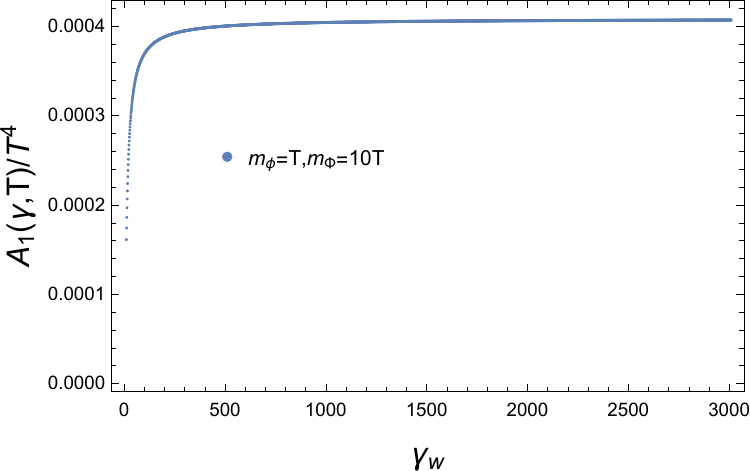} 
    \includegraphics[width=0.45\textwidth]{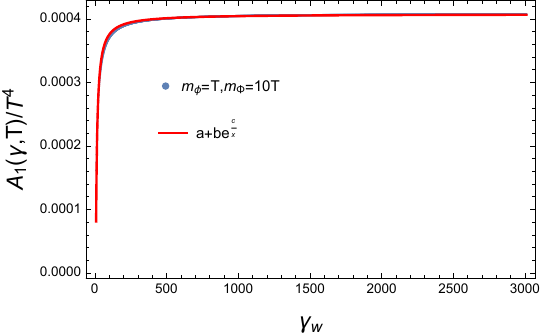}
    \caption{The left panel is the numerical result of $A_1(\gamma,T)$ as a function of Lorentz factor of bubble wall $\gamma_w$; The right panel is the fitting result of $A_1(\gamma,T)$ by quadratic function $a+b \exp(c/\gamma_w)$ with the fitting parameter is $a=-0.142253,b=0.142661$ and $c=-0.022931$.}
    \label{presdiagphi}
\end{figure}

\begin{figure}[htbp] 
    \centering 
    \includegraphics[width=0.45\textwidth]{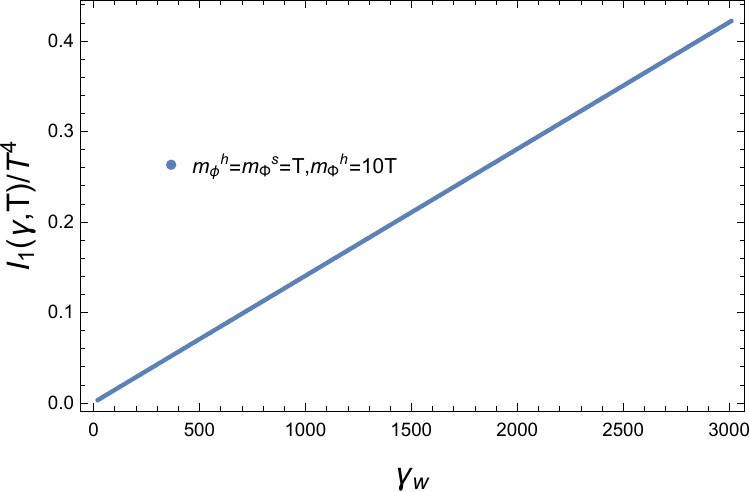} 
    \includegraphics[width=0.45\textwidth]{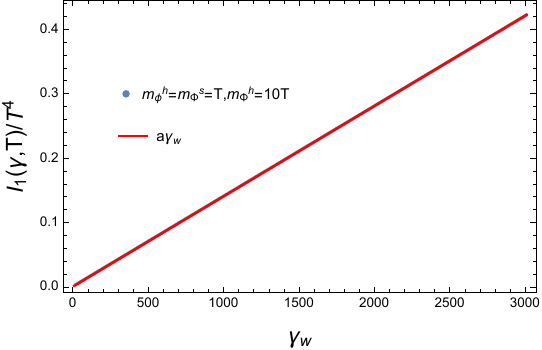}
    \caption{The Left panel is the numerical result of $I_1(\gamma,T)$ as a function of Lorentz factor of bubble wall $\gamma_w$; The Right panel is the fitting result of $I_1(\gamma,T)$ by liner function $a\gamma_w$ with the best fitting parameter is $a=0.00014$.}
    \label{presdiagphi}
\end{figure}

\begin{figure}[htbp] 
    \centering 
    \includegraphics[width=0.45\textwidth]{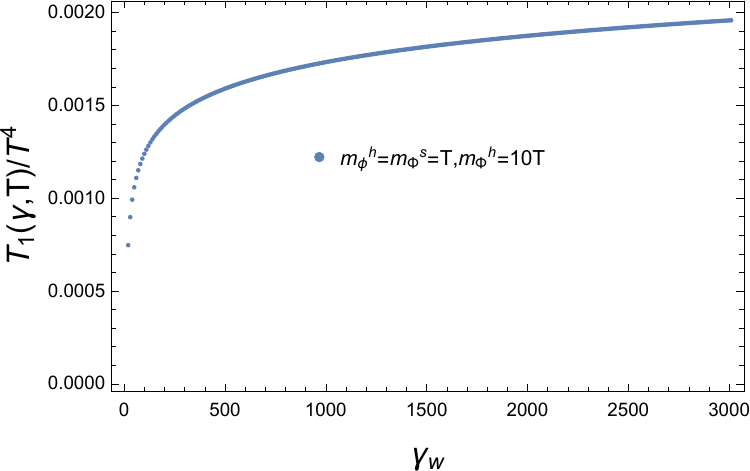} 
    \includegraphics[width=0.45\textwidth]{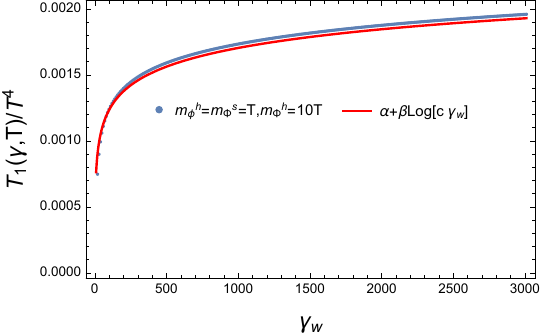}
    \caption{The Left panel is the numerical result of $T_1(\gamma,T)$ as a function of Lorentz factor of bubble wall $\gamma_w$; The Right panel is the fitting result of $T_1(\gamma,T)$ by quadratic function $\alpha+\beta\log(c \gamma_w)$ with the best fitting parameter is $\alpha=0.000801$, $\beta=0.000204$ and $c=0.079789$.}
    \label{presdiagphi}
\end{figure}
\begin{figure}[htbp] 
    \centering
    \includegraphics[width=0.45\textwidth]{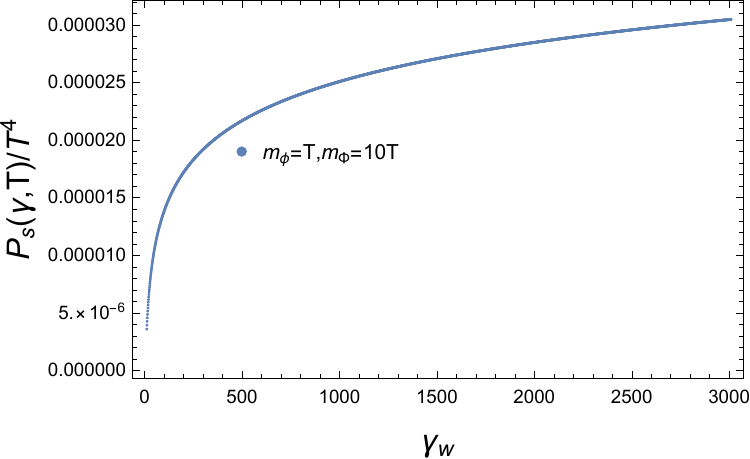} 
    \includegraphics[width=0.45\textwidth]{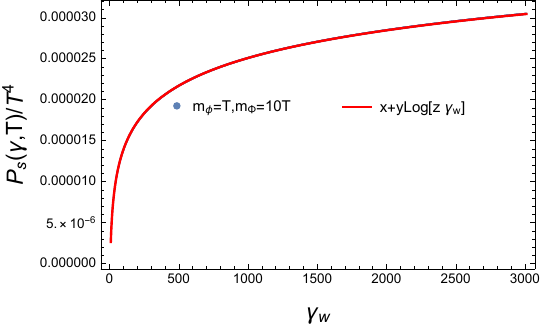} 
    \caption{The Left panel is the numerical result of $P_s(\gamma,T)$ as a function of Lorentz factor of bubble wall $\gamma_w$; The Right panel is the fitting result of $P_s(\gamma,T)$ by quadratic function $x+y\log(z \gamma_w)$ with the best fitting parameter is $x=0.0000252544$, $y=4.87588\times10^{-6}$ and $c=0.000963757$.}
    \label{presdiagphi}
\end{figure}

One can find the friction pressure from the $\phi^2\rightarrow\Phi^2$ process in the phase transition containing three $\gamma_w$ dependent functions $A_1$, $I_1$ and $T_1$. However, $R_1$ is completely similar to $A_1$ in Eq.\eqref{G5}. To figure out which part of the friction pressure dominates when the bubble velocity is large enough, we can plot them as a function of $\gamma_w$ using numerical computation. The result is shown in Fig.\ref{presdiagphi}. From the left panel of Fig.\ref{presdiagphi}, we can find that $A_1, T_1$ grow much slower than $I_1$. The numerical result for $I_1$ can be fit by a linear function of $\gamma_w$, as shown in the medium right panel of Fig.\ref{presdiagphi}, 
\begin{equation}
    \frac{F^{fric}_{\phi^2\rightarrow\Phi^2}}{A}\propto \gamma_w T^4.
\end{equation}
which means the $\phi^2\rightarrow\Phi^2$ friction pressure contributed by $I_1$ increases rapidly with the increase of $\gamma_w$ and would be the dominant part of the friction force if the bubble wall velocity is large enough. 


\vspace{-.3cm}

\bibliographystyle{unsrt}  
\bibliography{ref}

\end{document}